\documentclass[11pt]{article}

\usepackage[utf8]{inputenc}
\usepackage{amsmath, amssymb, amsfonts, graphicx}
\usepackage[colorinlistoftodos]{todonotes}
\usepackage[colorlinks=true, allcolors=blue]{hyperref}
\usepackage{listings, color}
\usepackage{enumitem}
\usepackage{float}
\usepackage{textcomp}
\usepackage{authblk}
\usepackage[numbers,super]{natbib}
\usepackage{hyperref}
\usepackage{fullpage}
\usepackage{multirow}
\usepackage{url}

\begin{document}

\title{A machine learning approach for efficient multi-dimensional integration}
\author[1\footnote{boram@lanl.gov}]{Boram Yoon}
\affil[1]{CCS-7, Applied Computer Science, Los Alamos National Laboratory,\protect\\Los Alamos, NM 87545, USA}
\date{}

\maketitle

\begin{abstract}
We propose a novel multi-dimensional integration algorithm using a machine learning (ML) technique. After training a ML regression model to mimic a target integrand, the regression model is used to evaluate an approximation of the integral. Then, the difference between the approximation and the true answer is calculated to correct the bias in the approximation of the integral induced by a ML prediction error. Because of the bias correction, the final estimate of the integral is unbiased and has a statistically correct error estimation. The performance of the proposed algorithm is demonstrated on six different types of integrands at various dimensions and integrand difficulties. The results show that, for the same total number of integrand evaluations, the new algorithm provides integral estimates with more than an order of magnitude smaller uncertainties than those of the VEGAS algorithm in most of the test cases.

\end{abstract}

\section{Introduction}\label{sec:intro}
Monte Carlo integration is a numerical method evaluating the integral of an integrand over a finite region using random sampling. As a consequence of the random sampling, in contrast to deterministic methods, the result of the Monte Carlo integration is an estimate of the true value that comes with a statistical uncertainty. For higher-dimensional integral problems, the Monte Carlo integration methods provide smaller uncertainties than deterministic methods, such as the trapezoidal rule~\cite{10.5555/1403886}, for a given number of integrand evaluations.
The most widely used strategies reducing the variance of the Monte Carlo integration estimate are importance sampling and stratified sampling. Two of the most popular algorithms implementing these strategies are VEGAS~\cite{PETERLEPAGE1978192, Lepage:123074} and MISER~\cite{10.5555/1403886}. Recently, an idea utilizing generative machine learning (ML) models to perform the importance sampling is also proposed \cite{Bendavid:2017zhk}.

In this paper, we present a novel algorithm numerically evaluating multi-dimensional integrals exploiting the efficient interpolation ability of ML regression algorithms. The new algorithm involves computations for the training of and prediction with the ML algorithms, in addition to the evaluation of the integrand. Assuming that evaluation of the integrand is computationally much more expensive than the ML calculations, throughout the paper, we focus on reducing the variance of an integral estimate for a given number of integrand evaluations.

\section{Method}
Suppose that we have a ML regression model $\tilde{f}(\mathbf{x})$ that approximates a multidimensional function $f(\mathbf{x}) \approx \tilde{f}(\mathbf{x})\equiv \tilde{y}$. Here $\mathbf{x}$ is the input vector of the regression model, which is called the \emph{independent variable}, and $\tilde{y}$ is the output of the regression model, which is called the \emph{dependent variable}. The integral of $f(\mathbf{x})$ over an integration region $\Omega \in \mathbb{R}^D$ can be split into two integrals
\begin{align}
  I &= \int_\Omega f(\mathbf{x}) d\mathbf{x}
  \label{eq:I}
  \\
    &= \underbrace{\int_\Omega \tilde{f}(\mathbf{x}) d\mathbf{x}}_{I_{\textrm{(a)}}}
      +\underbrace{\int_\Omega \left(f(\mathbf{x}) - \tilde{f}(\mathbf{x})\right) d\mathbf{x}}_{I_{\textrm{(b)}}}\,.
  \label{eq:split}
\end{align}
Here the $I_\textrm{(a)}$ is an integration of the ML regression model, which does not require an evaluation of $f(\mathbf{x})$ to calculate. Depending on the regression model, the integral $I_\textrm{(a)}$ may be calculated analytically or numerically. Since $\tilde{f}(\mathbf{x})$ approximates $f(\mathbf{x})$, the $I_\textrm{(a)}$ provides an approximation of the target integral $I$. 
The $I_\textrm{(b)}$ in Eq.~\eqref{eq:split} provides a correction to the bias of the approximation in $I_\textrm{(a)}$. The integral can be evaluated using a numerical method, such as the Monte Carlo integration. In the Monte Carlo integration, variance of the $I_\textrm{(b)}$ is proportional to
\begin{align}
  \textrm{Var}\big[f(\mathbf{X}) - \tilde{f}(\mathbf{X})\big] 
  &= \textrm{Var}\big[f(\mathbf{X})\big] + \textrm{Var}\big[\tilde{f}(\mathbf{X})\big] - 2\textrm{Cov}\big[f(\mathbf{X}), \tilde{f}(\mathbf{X})\big] \\
  &\approx 2 \textrm{Var}\big[f(\mathbf{X})\big] \left( 1 - \textrm{Corr}\big[f(\mathbf{X}), \tilde{f}(\mathbf{X})\big] \right)\,,
\end{align}
where the second line assumes a good ML regression model that gives $\textrm{Var}\big[\tilde{f}(\mathbf{X})\big] \approx \textrm{Var}\big[f(\mathbf{X})\big]$. Therefore, the $I_\textrm{(b)}$ can be estimated precisely with a small number of Monte Carlo samples when correlation between $f(\mathbf{X})$ and $\tilde{f}(\mathbf{X})$ is high. The total uncertainty of the integral, $\sigma_I$, can be calculated by combining the errors of the two terms: $\sigma_I^2 = \sigma_{I_{\textrm{(a)}}}^2 + \sigma_{I_{\textrm{(b)}}}^2$. The idea replacing an observable by its approximation with a proper correction term was proposed in the field of lattice quantum chromodynamics~\cite{Bali:2009hu,Blum:2012uh,Yoon:2018krb}. In this paper, we generalize the idea for multi-dimensional integral problems using ML regression algorithms. Now the question is how to obtain a good ML regression model that closely approximates the integrand.

\subsection*{Machine learning models}
In this paper, we examine three regression algorithms: Multi-layer Perceptron (MLP), Gradient Boosting Decision Tree (GBDT), and Gaussian Processes (GP). MLP is a feedforward artificial neural network that produces outputs from inputs based on multiple layers of perceptrons~\cite{10.5555/235222}. The model is flexibly applicable to various kinds of data and scales well up to a large number of data. GBDT is a sequence of shallow decision trees such that each successive decision tree compensates for the prediction error of its predecessor~\cite{breiman1984classification, friedman2001}. The model provides a good regression performance with no complicated tuning of hyperparameters and pre-processing of training data. An integration of the GBDT regression models can be calculated analytically because the model is simply a set of intervals of input variables and their output values~\cite{Bendavid:2017zhk}. GP regression is a nonparametric model that finds an optimal covariance kernel function explaining training data~\cite{3569}. The model is good at interpolating the observations and works well with a small data set. Analytic integrability of the regression model depends on kernel choice. For example, in the case of the Radial Basis Function (RBF) kernel, which is one of the most popular kernels in GP, prediction of an input vector is given by the dot product of a Gaussian function of the input vector and a constant vector, as described in Eq.~\eqref{eq:gpr_nrt}, so its analytic integration is given by error functions.

\subsection*{Training data}
Building a ML regression model approximating $f(\mathbf{x})$ requires training samples of $\{(\mathbf{x}_i, f(\mathbf{x}_i))\}_{i=1}^{N_{\text{train}}}$. To minimize the prediction error for a given number of training data, $N_{\text{train}}$, it is essential to collect the training samples near the peaks of the function (importance sampling) and where the function changes rapidly (stratified sampling). Such training data can be sampled by utilizing conventional numerical integration algorithms, such as the VEGAS, which includes efficient sampling algorithms based on the importance sampling and stratified sampling. When the peaks of the function are localized, the training samples obtained using VEGAS build a much more accurate ML regression model than those from a uniform sampling method.

\subsection*{Data scaling}
Many ML regression algorithms benefit from scaling the dependent variable. Especially when the dependent variable varies by orders of magnitude within the range of interest, which is a typical situation in difficult multi-dimensional integral problems, the data scaling plays a crucial role in obtaining a good regression performance. The most widely used scaling algorithms are min-max scaling and standardization:
\begin{align}
  y' = \frac{y-\text{min}(y)}{\text{max}(y)-\text{min}(y)} \quad \textrm{[Min-max scaling]}, \qquad\qquad
  y' = \frac{y - \overline{y}}{\sigma_y} \quad \textrm{[Standardization]}\,,
\end{align}
where $y'$ is the scaled variable, $\text{min}(y)$ and $\text{max}(y)$ are the minimum and maximum of $y$, $\overline{y}$ is the average of $y$, and $\sigma_y$ is the standard deviation of $y$. For the data with large scale variation, however, these scaling methods are dominated by the data of large magnitude and lose sensitivity to the data of small values.

To avoid the scale issue, we use the nth-root scaling defined as
\begin{align}
  y' = \textrm{sgn}(y) \cdot |y|^{1/n}\,,
  \label{eq:nth-root}
\end{align}
where $\textrm{sgn}(y)$ is the sign of $y$, and $n$ is a positive integer. This is a strictly monotonic transformation whose inverse is $y = \textrm{sgn}(y') \cdot|y'|^n$.
%The nth-root transformation shares some similarity with the power transformation \cite{10.1093/biomet/87.4.954, Box.Cox1964}. However, the power transformation is focused on restoring the normality of data distribution, while the nth-root transformation is focused on reducing the scaling of data.
The optimal value of $n$ can be obtained using the training data. Taking a small portion (e.g. 10--50\%) of the training data as a validation dataset, one can train a regression model on the remaining training samples with various choices of $n$ and find the optimal value of $n$ that gives the minimum prediction error on the validation dataset. Once the $n$ is determined, a final regression model can be obtained using the full training data. 

This nth-root scaling plays a crucial role in building a good regression model for most of the integral problems. In this study, we standardize the $y$ after the nth-root scaling to maximize the regression performance.

\subsection*{Evaluation of $I_{\textrm{(a)}}$ and $I_{\textrm{(b)}}$}
When $n=1$, $I_{\textrm{(a)}}$ of Eq.~\eqref{eq:split} can be calculated analytically for certain regression algorithms. When $n>1$, however, the ML predictions should be processed by the inverse of the nth-root transformation, so the analytic integral becomes complicated. For example, GP regression with a RBF kernel for a nth-root scaled data can be written as
\begin{align}
  \tilde{f}'(\mathbf{x})
  = \sum_{i=1}^{N_{\text{train}}} \alpha_i \exp\left[-\frac{1}{2l^2}  || \mathbf{x} - \mathbf{x}_i ||^2\right]\,,
  \label{eq:gpr_nrt}
\end{align}
where $||\mathbf{x}||$ is the Euclidean norm of $\mathbf{x}$, $\mathbf{x}_i$ are the training data, and $\alpha_i$ and $l$ are constants that are determined from the training. To obtain the prediction of the integrand $\tilde{f}(\mathbf{x})$, the GP regression needs to be transformed as $\tilde{f}(\mathbf{x}) = \textrm{sgn}(\tilde{f}'(\mathbf{x})) |\tilde{f}'(\mathbf{x})|^n$. For a positive integrand, the power of $n$ of Eq.~\eqref{eq:gpr_nrt} can be expanded analytically, but the number of terms is large for large $N$ and $n$.

For simplicity, we use a numerical method, the VEGAS algorithm, to evaluate $I_{\textrm{(a)}}$ and $I_{\textrm{(b)}}$. Since the peaks of the $f(\mathbf{x})$ are flattened by subtracting the $\tilde{f}(\mathbf{x})$ in the integrand of $I_{\textrm{(b)}}$, a simple Monte Carlo integration works well for $I_{\textrm{(b)}}$. However, the VEGAS outperforms the simple Monte Carlo integration when the regression is not accurate enough.

\section{Numerical Experiments}
In this section, we present numerical experiments of the proposed integration algorithm using ML. The precisions of the integral estimates are compared with those of the VEGAS algorithm, which is one of the best performing algorithms on the market~\cite{Hahn:2004fe}, at a similar number of integrand evaluations.

\subsection*{Test integrands}
In order to test the performance of the numerical integration, we use the six families of the integrands proposed in Ref.~\cite{Genz1987}:
\begin{equation}
\label{eq:families}
\begin{array}{ll}
  f_1(\mathbf{x}) = \cos(2\pi w_1 + \mathbf{c}\cdot\mathbf{x}) & \text{[Oscillatory]}\,, \\[1ex]
  f_2(\mathbf{x}) = \prod\limits_{i = 1}^{D}
    \dfrac 1{c_i^{-2} + (x_i - w_i)^2} & \text{[Product peak]}\,, \\[2ex]
  f_3(\mathbf{x}) = \dfrac 1{(1 + \mathbf{c}\cdot\mathbf{x})^{D + 1}} & \text{[Corner peak]}\,, \\[3ex]
  f_4(\mathbf{x}) = \exp(-\sum_{i=1}^D c_i^2 (x_i - w_i)^2) & \text{[Gaussian]}\,, \\[3ex]
  f_5(\mathbf{x}) = \exp(-\mathbf{c}\cdot |\mathbf{x} - \mathbf{w}|) & \text{[$C^0$-function]}\,, \\[1ex]
  f_6(\mathbf{x}) = \begin{cases}
    0 & \text{if }x_1 > w_1 \textrm{ or } x_2 > w_2 \,, \\
    \exp(\mathbf{c}\cdot\mathbf{x}) & \textrm{otherwise}.
  \end{cases} & \text{[Discontinuous]}\,.
\end{array}
\end{equation}
Here, $D$ is the dimension of $\mathbf{x}$, and $w_i \in [0, 1)$ is the parameter that is supposed to shift the peaks of the integrand without changing the difficulty of the integral problem. One exception is $f_6(\mathbf{x})$, as the small value of $w_1$ or $w_2$ makes the function to be localized in small region and makes the integral problem difficult. To avoid the unwanted effect, we restrict $w_i \in [0.1, 0.9)$ for $f_6(\mathbf{x})$. $c_i$ is a positive parameter that controls the difficulty of the integral. In general, increasing the value of $c_i$ increases the difficulty of the integral problem. To fix the difficulty of the integral, we randomly choose $c_i$ from a uniform distribution in $[0, 1)$ and renormalize the vector by multiplying a constant factor so that $||\mathbf{c}||_1 = \sum_i |c_i|$ becomes the target constant. In this study, we carry out the integration for 36 different random choices of $\mathbf{w}$ and $\mathbf{c}$ and take average performance. To fix the integration difficulty, we normalize $\mathbf{c}$ to three different values of $||\mathbf{c}||_1 = 1, 3$, and $8$. Integration is performed in a $D$-dimensional unit hypercube, and the results are compared at three different dimensions of $D=5, 8$, and $10$.

\subsection*{ML regression algorithms and hyperparameters}
For the implementation of the MLP, GP, and GBDT regression algorithms, we use the scikit-learn python library \cite{scikit-learn}. For MLP, four hidden layers of 128, 128, 128, and 16 neurons with rectified linear unit (ReLU) activation functions are used. Training is performed using Adam optimization algorithm\cite{kingma2014method} with the learning rate of $10^{-4}$. Training updates are performed until there is no decrease of the validation score with a tolerance of $10^{-6}$ for 20 epochs with 10\% of validation fraction. For GP, RBF with a constant kernel is used, and length scale and constant are determined using L-BFGS-B optimizer\cite{53712fe04a3448cfb8598b14afab59b3}. For GBDT, we use 1000 weak estimators with a learning rate of $0.01$ and a subsampling ratio of $0.3$. The maximum depth of each decision tree is limited to 4. Note that here we use a relatively large number of estimators with a small subsampling ratio so that the regression output becomes a smooth function in $\mathbf{x}$. In this proof-of-principal study, we did not explore extensive phase space of the hyperparameters but take those generally considered to be a reasonable choice. 

The powers of the nth-root scaling $n \in [1,50]$ for MLP and GP regressions are determined by using 20\% of training data as a validation dataset. The performance of the GBDT algorithm is not very sensitive to the scaling but $n>1$ gives better performance than the $n=1$ case. So, we use a fixed number $n=3$ for GBDT regression.

\subsection*{VEGAS setup}
For the VEGAS numerical integration, we use Lepage's VEGAS python library
\cite{peter_lepage_2020_3897199}. The library has two damping parameters: $\alpha$ and $\beta$. The parameter $\alpha$ controls the remapping of the integration variables in the VEGAS adaptation. A smaller value gives the slower grid adaption for a conservative estimate. Here, we use $\alpha=0.5$, which is the default value of the library, for most of the calculations. One exception is the discontinuous integrand family $f_6(\mathbf{x})$, which is more difficult to evaluate than other integrand families and requires a large number of samples per iteration or slow grid adaptation to converge to the exact integral solution. To make the VEGAS integral stable, we use $\alpha=0.2$ for $f_6(\mathbf{x})$. The parameter $\beta$ controls the redistribution of integrand evaluations in the stratified sampling. $\beta=1$ is the theoretically optimal value, and $\beta=0$ means no redistribution. Here, we use $\beta=0.75$, which is the default value of the library.

Another important parameters are the number of iterations for the VEGAS grid adaptation ($N_{\text{itn}}$) and the approximated number of integrand evaluations per iteration ($N_{\text{eval}}$). These parameters are set differently for different VEGAS tasks performed in this study: (1) calculation of the target integral $I$ in Eq.~\eqref{eq:I} for a comparison, (2) sampling the training data, (3) calculation of $I_{\textrm{(b)}}$ in Eq.~\eqref{eq:split}, and (4) calculation of $I_{\textrm{(a)}}$ in Eq.~\eqref{eq:split}. 

\begin{itemize}
\item In task (1), we use $N_{\text{itn}}=20$ at two different values of $N_{\text{eval}}=500$, and $1000$. When $\beta=0$, the total number of integrand evaluations will be $N_{\text{itn}}\times N_{\text{eval}}$. Because of the redistribution that happens when $\beta>0$, however, the total number of integrand evaluations for this task drops to around $N \approx N_{\text{itn}}\times N_{\text{eval}}/2$ for the test functions used in this study. 

\item In task (2), we use $N_{\text{itn}}=10$ for the most of the integrand families with the same $N_{\text{eval}}$ values used in task (1). In this task, all the integrand calls, $\big(\mathbf{x}, f(\mathbf{x})\big)$, are collected as the ML training data. The total number of integrand evaluations in this task is $N_{\text{train}} \approx N/2$. Two exceptions are $f_3(\mathbf{x})$, where we use $N_{\text{itn}}=14$, and $f_6(\mathbf{x})$, where we use $N_{\text{itn}}=6$, which are the choices that stabilize the VEGAS integration of task (3). The choice of $N_{\text{itn}}$ determines the ratio of the number of integrand evaluations for the training ($N_{\textrm{train}})$ and bias correction ($N_{\textrm{crxn}}$). The parameter can be tuned for a given problem so that it minimizes the integral uncertainty. In this proof-of-principal study, however, we take $N_{\text{itn}}=10$, which makes the ratio $N_{\textrm{train}}:N_{\textrm{crxn}} \approx 1:1$, as our default value and change it only when we find an instability in the VEGAS integration of task (3).

\item In task (3), our target total number of integrand evaluations is $N - N_{\text{train}}$, so that the total number of integrand evaluations in the ML integrator is the same as that of the VEGAS integration in task (1). For that, we set $N_{\text{eval}} = (N - N_{\text{train}})/5$ with $\alpha=0.5$ and stop the VEGAS iteration when the accumulated number of integrand evaluations, $N_{\text{crxn}}$, is close to $N - N_{\text{train}}$. %Here, a small $N_{\text{eval}}$ is used to stop the iteration as close as possible to the target number of integrand evaluations, and a small  $\alpha$ is used to stabilize the VEGAS integration for small $N_{\text{eval}}$.

\item In task (4), we use $N_{\text{itn}}=30$ and set $N_{\text{eval}}$ to those used in task (1) multiplied by a factor of 1000. We stop the VEGAS iteration when the error of $I_{\textrm{(a)}}$ becomes smaller than 20\% of the error of $I_{\textrm{(b)}}$.
\end{itemize}

VEGAS integral estimates are obtained by taking a weighted average of the estimates from each VEGAS iteration. Whenever the p-value of the weighted average is smaller than 0.05, we discard the results and rerun the VEGAS integration with a different random seed.

\subsection*{Results}

\begin{table}[tb]
    \centering
    \begin{tabular}{ccc|rrrrrr}
\hline\hline
\multicolumn{3}{c|}{Integrand family} & \multicolumn{1}{c}{1} & \multicolumn{1}{c}{2} & \multicolumn{1}{c}{3} & \multicolumn{1}{c}{4} & \multicolumn{1}{c}{5} & \multicolumn{1}{c}{6} \\
\multicolumn{3}{c|}{(ML Algorithm)} & \multicolumn{1}{c}{(GP)} & \multicolumn{1}{c}{(GP)} & \multicolumn{1}{c}{(GP)} & \multicolumn{1}{c}{(GP)} & \multicolumn{1}{c}{(MLP)} & \multicolumn{1}{c}{(GBDT)} \\
\hline
N & D & $||\mathbf{c}||_1 $ &&&&&&\\
\multirow{9}{*}{$\approx$5000} &
\multirow{3}{*}{5} &
   1.0 &  407(29) &  6181(319) &  329(4)  &   6223(258) &  14.1(6) &  10.0(5) \\
&& 3.0 &   197(2) &   1321(67) &  211(98) &    1151(47) &   5.5(2) &   6.5(4) \\
&& 8.0 &   181(1) &      47(8) &  138(5)  &      346(9) &   3.1(1) &   3.1(2) \\\cline{2-9}
& \multirow{3}{*}{8} &
   1.0 &  352(28) &  8292(262) &  207(3)  &   8918(177) &   8.9(3) &   8.6(6) \\
&& 3.0 &   141(2) &  1219(100) &  130(1)  &    1606(46) &   3.6(1) &   4.3(2) \\
&& 8.0 &   107(3) &      20(2) &   57(3)  &      319(7) &   1.6(1) &   2.3(1) \\\cline{2-9}
& \multirow{3}{*}{10} &
   1.0 &  345(29) &  8533(250) &  178(2)  &   8941(189) &   6.5(2) &   8.2(4) \\
&& 3.0 &   119(1) &   972(106) &   99(1)  &    2028(56) &   2.5(1) &   4.3(2) \\
&& 8.0 &    54(2) &      17(2) &   35(1)  &     386(10) &   1.2(1) &   2.0(1) \\
\hline
\multirow{9}{*}{$\approx$10000} &
\multirow{3}{*}{5} &
   1.0 &  357(24) &  5770(311) &  299(3)  &   5497(235) &  16.7(5) &  14.5(7) \\
&& 3.0 &   175(2) &   1241(44) &  200(92) &    1084(44) &   7.9(3) &   7.5(3) \\
&& 8.0 &   157(5) &    121(20) &  144(6)  &      279(8) &   4.0(1) &   3.6(2) \\\cline{2-9}
& \multirow{3}{*}{8} &
   1.0 &  359(28) &  9130(307) &  215(3)  &   9649(203) &  16.7(5) &  11.0(6) \\
&& 3.0 &   147(2) &   1592(89) &  142(1)  &    1698(49) &   6.6(2) &   5.0(2) \\
&& 8.0 &   132(2) &  40(5)     &   97(3)  &      320(8) &   3.1(1) &   2.7(1) \\\cline{2-9}
& \multirow{3}{*}{10} &
   1.0 &  339(27) &  9092(218) &  181(2)  &   9535(118) &  13.0(5) &   9.8(6) \\
&& 3.0 &   118(1) &  1679(111) &  116(1)  &    1976(47) &   5.1(2) &   4.7(3) \\
&& 8.0 &    89(3) &  36(4)     &   57(2)  &      372(9) &   2.4(1) &   2.2(1) \\
\hline\hline
    \end{tabular}
    \caption{Precision gain of the proposed algorithm over VEGAS, defined in Eq.~\protect\eqref{eq:gain}, at two different number of integrand evaluations ($N$) for three dimensions ($D$) and three integrand difficulties ($||\mathbf{c}||_1$) on the six integrand families listed in Eq.\protect~\eqref{eq:families}. The results are averaged over 36 random samples. The numbers in the parentheses are the standard deviation of the mean.}
    \label{tab:gain}
\end{table}

\begin{figure}
    \centering
    \includegraphics[width=0.47\textwidth]{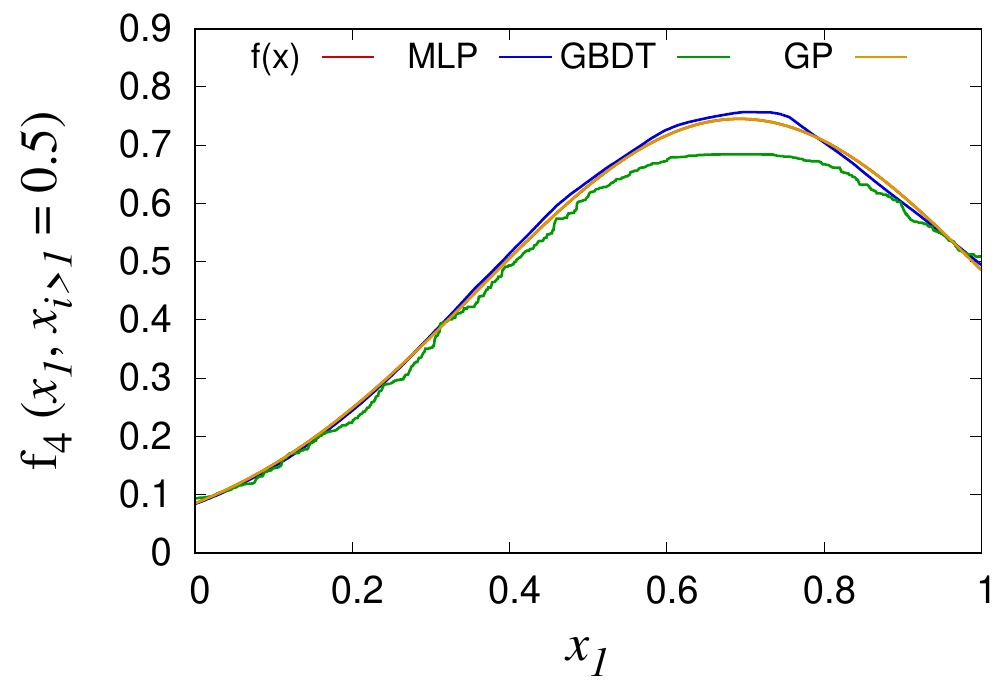}\qquad
    \includegraphics[width=0.47\textwidth]{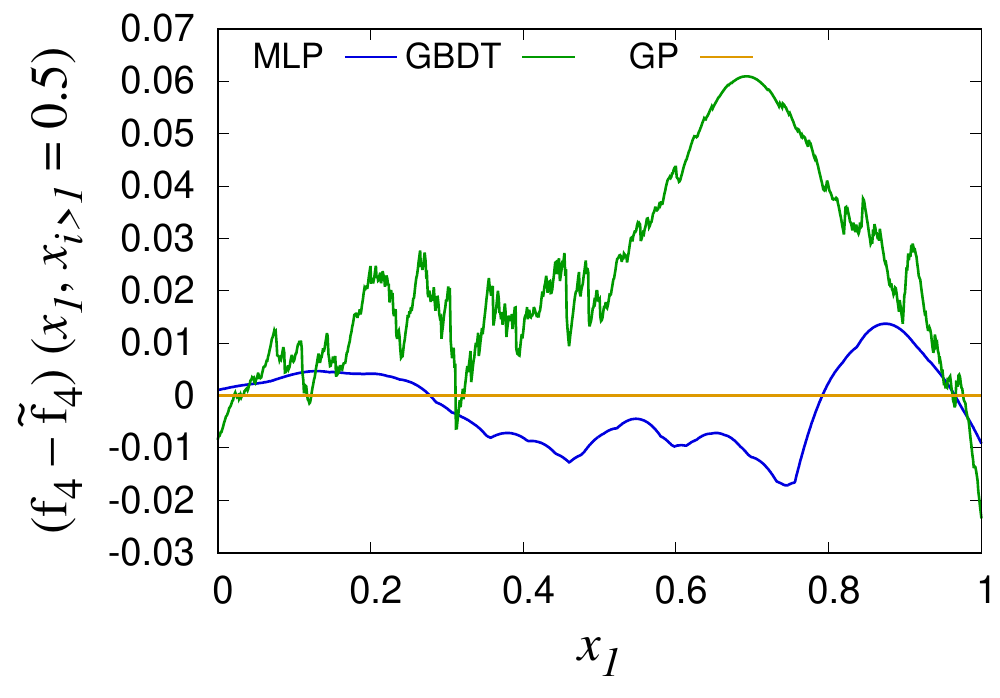}\\
    \includegraphics[width=0.47\textwidth]{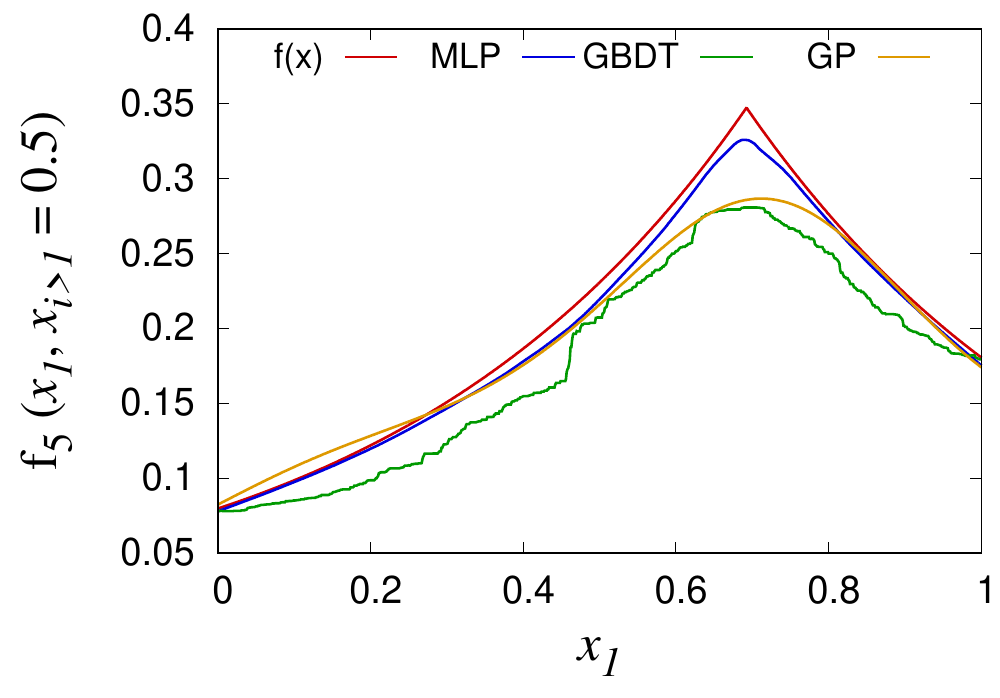}\qquad
    \includegraphics[width=0.47\textwidth]{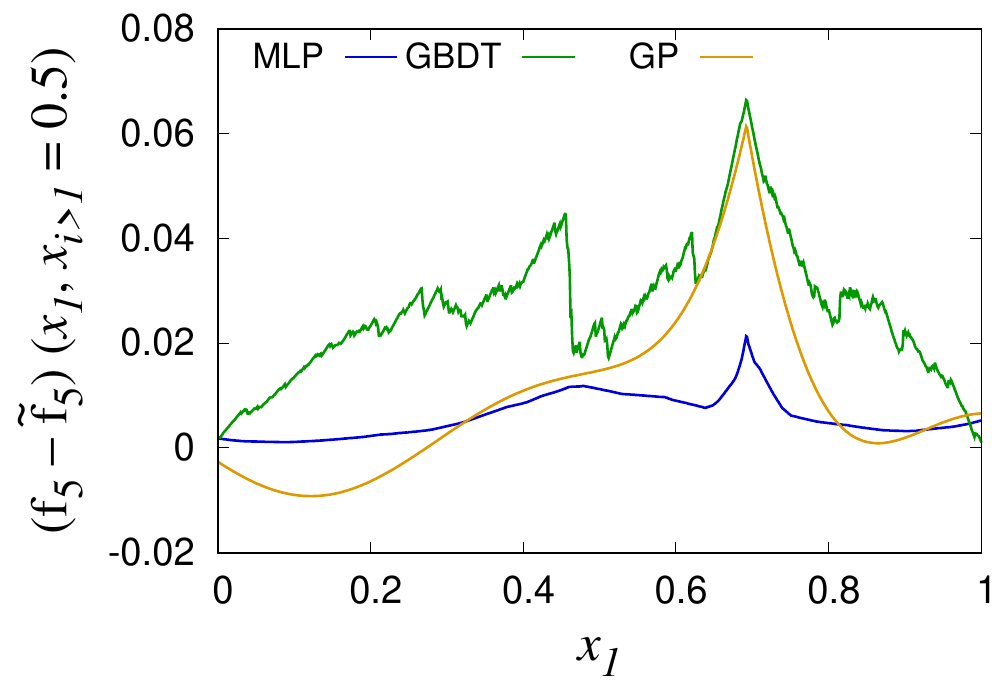}\\
    \includegraphics[width=0.47\textwidth]{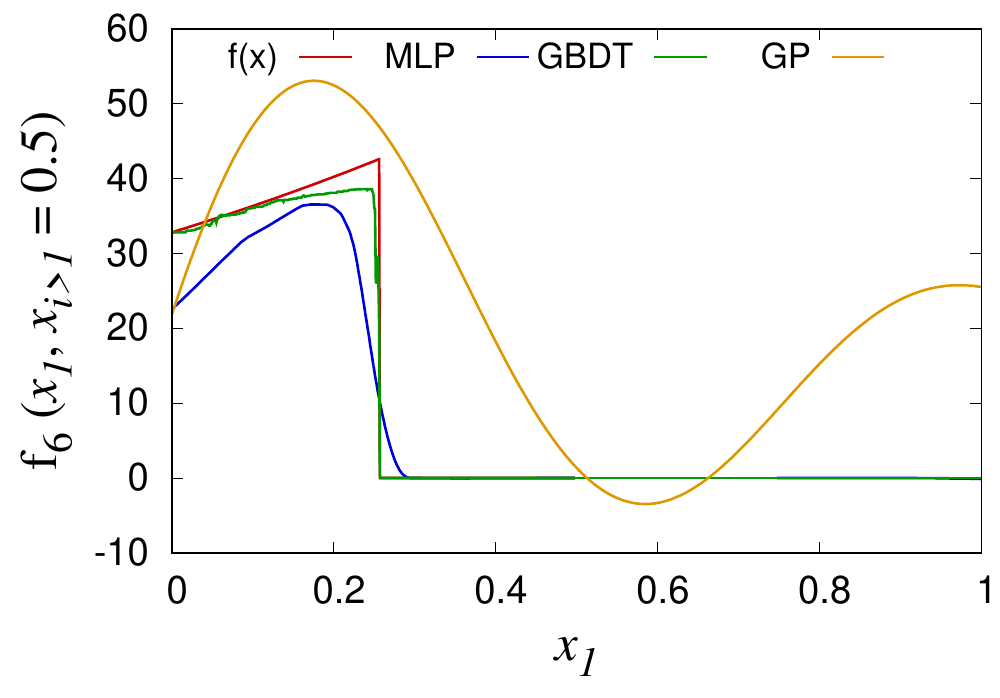}\qquad
    \includegraphics[width=0.47\textwidth]{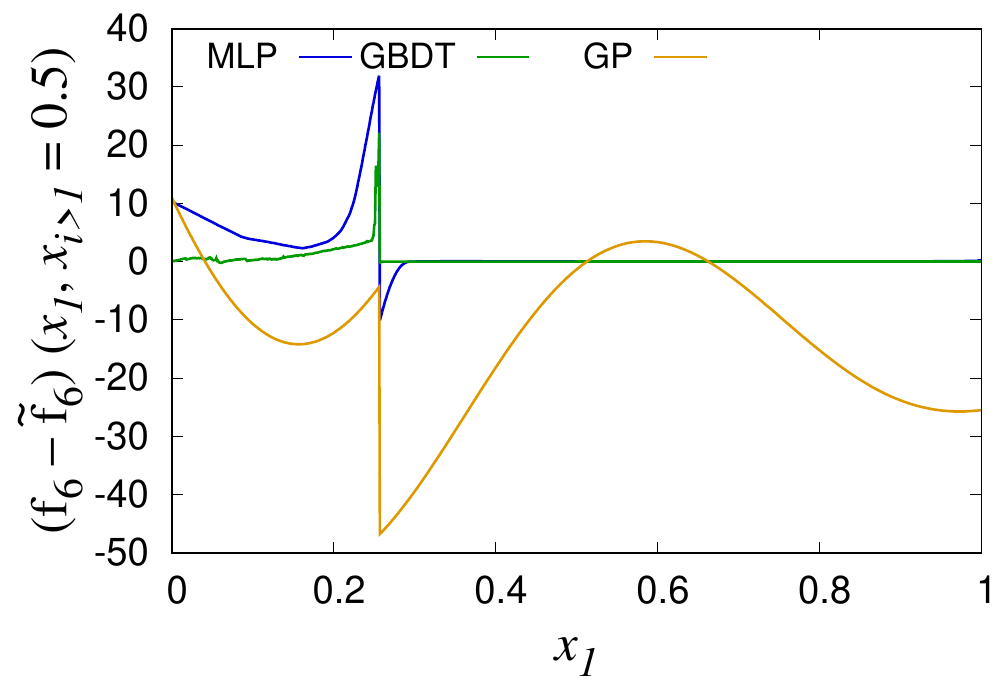}\\
    \caption{Integrands $f_i(\mathbf{x})$ and their ML predictions $\tilde{f}_i(\mathbf{x})$ (left), and the prediction errors ${f_i(\mathbf{x}) - \tilde{f}_i(\mathbf{x})}$ (right) for $N\approx 5000$, D=8, and $||\mathbf{c}||_1=8.0$. Those are plotted as a function of $x_1$, while rest of the $\mathbf{x}$ are fixed to $x_{i=2,3,\ldots,8}=0.5$.}
    \label{fig:funcs}
\end{figure}

Table~\ref{tab:gain} shows the precision gain of the proposed integration algorithm over VEGAS,
\begin{align}
    \textrm{Gain} = \frac{\sigma_I \textrm{ of VEGAS}}{\sigma_I \textrm{ of ML Integrator}}\,.
    \label{eq:gain}
\end{align}
Total number of integrand evaluations of the ML integrator ($N_{\textrm{train}}+N_{\textrm{crxn}}$) is similar to that of the VEGAS integration ($N$). The full list of $N$, $N_{\textrm{train}}$, $N_{\textrm{crxn}}$, and precision of the integral algorithms are given in Appendix~\ref{sec:appendix-res}.

The best performing ML algorithms are GP for the integrand families 1--4, MLP for the integrand family 5, and GBDT for the integrand family 6. Fig.~\ref{fig:funcs} clearly explains these results: GP with a RBF kernel shows very good performance in describing smooth functions but fails in $C^0$ and discontinuous functions. MLP shows mediocre performance for the all functional forms, and GBDT, which is a combination of the discrete decision trees, outperforms the MLP in describing the discontinuous integrands.

For all test cases, the ML integrator performs better than VEGAS. The gain is higher when $D$ is smaller and when $||\mathbf{c}||_1$ is smaller. Also, the gain tends to be increased when $N$ is larger, which indicates a better scaling behavior than VEGAS. In case of the integrations with the GP regression algorithm, the $\sigma_I$ of the ML integrator is up to four orders of magnitude smaller than that of VEGAS. When GP is efficiently applied, the difference between the target integrand and its ML prediction is tiny, which makes the value and error of $I_{\textrm{(b)}}$ small. As a result, the final error is dominated by the error of $I_{\textrm{(a)}}$, which can be improved without increasing the number of $f(\mathbf{x})$ evaluations. As an example, below shows $I_{\textrm{(a)}}$ and $I_{\textrm{(b)}}$ of the integrands shown in Fig.~\ref{fig:funcs}:
\begin{align}
\begin{array}{lllll}
     & I             & = I_{\textrm{(a)}} & +I_{\textrm{(b)}} &\\
f_4: & 0.3195642(12) & = 0.3195644(12) &- 0.000000189(30) & [GP]\\
f_5: & 0.11582(10)   & = 0.115408(17)  &+ 0.00041(10)     & [MLP]\\
f_6: & 14.463(42)    & = 13.8934(51)   &+ 0.570(42)        & [GBDT]\\
\end{array}
\end{align}

Since the ML integrator uses VEGAS, it inherits the potential instability of the VEGAS for small $N_{\textrm{eval}}$, which introduces a systematic bias in the integration results. In general, the instability can be avoided by increasing $N$ or decreasing the value of $\alpha$; a detailed description of how to deal with the instability is given in Ref.~\cite{VEGAS_DOC}. For the ML integrator, the most fragile part is the integration for $I_{\textrm{(b)}} = \int(f - \tilde{f})d\mathbf{x}$. When such instability is observed, for a given $N$, one can increase the ratio of $N_{\textrm{train}}$ for a better prediction or increase the ratio of $N_{\textrm{crxn}}$ for a more stable integrand evaluation, depending on the integral problem. It is also important to use a ML regression algorithm that yields a smooth $\tilde{f}(\mathbf{x})$. As shown in the right column of Fig.~\ref{fig:funcs}, a non-smooth $\tilde{f}(\mathbf{x})$, such as the one from GBDT, makes $f(\mathbf{x})-\tilde{f}(\mathbf{x})$ highly oscillating and the integral difficult to evaluate. Among the three regression algorithms used in this study, we find that the smooth prediction of GP gives the most stable integration for $I_{\textrm{(b)}}$. To check the instability, we do not manually tune the integration parameters for each integral problem but use a general setting for most of the calculations. As a result, we could observe a few integral results deviating from the true answer by more than $4\sigma$, mostly in case the integral families 3 and 6. Since the number of such occurrences is small compared to the total number of random samples, the inclusion of these occurrences does not change the average results, so we did not exclude these results from our average estimation. The number of more than $4\sigma$ deviations for each integral problem is given in the tables in Appendix~\ref{sec:appendix-res}.

\section{Conclusion}
In this paper, we proposed a novel algorithm calculating multi-dimensional integrals using ML regression algorithms. In this algorithm, a ML regression model is trained to mimic the target integrand and is used to estimate an approximated integral. Any bias of the estimate induced by the ML prediction error is corrected by using a bias correction term, as described in Eq.~\eqref{eq:split}, so that the final integral result could have a statistically correct estimation of the uncertainty. Two essential prescriptions for obtaining a good the training efficiency are (1) collecting training samples using the VEGAS algorithm, and (2) scaling the training data using the nth-root scaling defined in Eq.~\eqref{eq:nth-root}.

The performance of the proposed ML integrator is compared with that of the VEGAS algorithm on six different integrand families listed in Eq.~\eqref{eq:families}. Three ML regression algorithms of MLP, GBDT, and GP are examined, and the best performing algorithm is selected for each integrand family. For all test cases, the ML integrator shows better performance than the VEGAS for a given total number of integrand evaluations. In most of the cases, the ML integrator is able to provide integration results with more than an order of magnitude smaller uncertainty than the VEGAS algorithm. The performance gain is presented in Table~\ref{tab:gain}. As a proof-of-principal study, we did not tune the ML and algorithm hyperparameters for each integral problem. By tuning the parameters tailored to a specific problem, one would be able to further reduce the integration error of the ML integrator.

We find that the performance and the stability of the proposed algorithm largely depend on the smoothness of the regression output. Developing a ML algorithm specifically targeting the ML integrator will be able to improve the performance and stability of the algorithm. One possible approach is to augment the training data by adding a small amount of noise to the training data~\cite{105415,An1996TheEO}, which could improve the smoothness of the MLP and GBDT models. We also find that the GP regression algorithm with a RBF kernel fails in describing $C^0$ and discontinuous functions because of the singular points in the integrands. For a given integrand with known such singular points, one would be able to build a combination of multiple GP models defined on each domain divided by the singular points for a better performance. It will be also promising to explorer different types of kernels~\cite{SCHULZ20181} or to develop a hybrid model of decision tree and GP that can be generically applicable for such integrands.

\section*{Acknowledgments}
Computations were carried out using Institutional Computing at Los Alamos National Laboratory. This work was supported by the U.S. Department of Energy, Office of Science, Office of High Energy Physics under Contract No. 89233218CNA000001, and the LANL LDRD program.

\appendix

\section{Tables of results}
\label{sec:appendix-res}
In Tables~\ref{tab:res5000-5}--\ref{tab:res10000-10}, we show the results of VEGAS and ML integrations algorithms.

%%%%%%%%%%%%%%%%%%%%%%%%%%%%%%%%%%%%%%%%%%
\begin{table}[htbp]
\footnotesize
    \centering
    \begin{tabular}{cc|cccccc}
\multicolumn{8}{c}{N$\approx$5000, D=5, $||\mathbf{c}||_1$=1.0}\\\hline\hline
\multicolumn{2}{c|}{Integrand family} & 1 & 2 & 3 & 4 & 5 & 6 \\\hline
\multicolumn{2}{c|}{$N$}&          5313(27) &          5514(13) &       5135.3(3.2) &          5520(15) &       5255.6(7.5) &          5869(30) \\
\multicolumn{2}{c|}{$N_{\textrm{train}}$}&          2623(14) &       2732.0(8.9) &       3590.5(2.8) &       2735.4(8.0) &       2546.9(4.8) &          1758(12) \\
\multicolumn{2}{c|}{$N_{\textrm{crxn}}$}&          2739(25) &          2845(19) &       1545.4(5.0) &          2824(14) &          2775(16) &          4118(50) \\
\hline
\multirow{4}{*}{$\sigma_I/|I|$}
& VEG & $5(3)\hspace{-0.3em}\times\hspace{-0.3em} 10^{-3}$ & $3.6(1)\hspace{-0.3em}\times\hspace{-0.3em} 10^{-4}$ & $8.8(1)\hspace{-0.3em}\times\hspace{-0.3em} 10^{-4}$ & $3.7(1)\hspace{-0.3em}\times\hspace{-0.3em} 10^{-4}$ & $5.04(8)\hspace{-0.3em}\times\hspace{-0.3em} 10^{-4}$ & $4.7(3)\hspace{-0.3em}\times\hspace{-0.3em} 10^{-3}$ \\
& MLP & $1.4(8)\hspace{-0.3em}\times\hspace{-0.3em} 10^{-4}$ & $9.2(3)\hspace{-0.3em}\times\hspace{-0.3em} 10^{-6}$ & $1.31(4)\hspace{-0.3em}\times\hspace{-0.3em} 10^{-4}$ & $8.9(2)\hspace{-0.3em}\times\hspace{-0.3em} 10^{-6}$ & $3.7(1)\hspace{-0.3em}\times\hspace{-0.3em} 10^{-5}$ & $4.6(4)\hspace{-0.3em}\times\hspace{-0.3em} 10^{-3}$ \\
& GBDT & $2(1)\hspace{-0.3em}\times\hspace{-0.3em} 10^{-3}$ & $1.68(3)\hspace{-0.3em}\times\hspace{-0.3em} 10^{-5}$ & $7.7(1)\hspace{-0.3em}\times\hspace{-0.3em} 10^{-4}$ & $1.72(4)\hspace{-0.3em}\times\hspace{-0.3em} 10^{-5}$ & $1.35(2)\hspace{-0.3em}\times\hspace{-0.3em} 10^{-4}$ & $4.8(3)\hspace{-0.3em}\times\hspace{-0.3em} 10^{-4}$ \\
& GP & $2(1)\hspace{-0.3em}\times\hspace{-0.3em} 10^{-5}$ & $6.6(4)\hspace{-0.3em}\times\hspace{-0.3em} 10^{-8}$ & $2.69(4)\hspace{-0.3em}\times\hspace{-0.3em} 10^{-6}$ & $6.3(3)\hspace{-0.3em}\times\hspace{-0.3em} 10^{-8}$ & $1.83(4)\hspace{-0.3em}\times\hspace{-0.3em} 10^{-4}$ & $2.4(3)\hspace{-0.3em}\times\hspace{-0.3em} 10^{-2}$ \\
\hline
\multirow{4}{*}{Gain}
& MLP &         37.9(3.4) &         41.2(1.5) &          6.91(18) &         42.2(1.3) &         14.06(57) &         1.177(65) \\
& GBDT &          5.06(41) &         22.10(86) &         1.166(34) &         21.83(94) &          3.77(10) &         10.03(51) \\
& GP &           407(29) &         6181(319) &        329.0(4.1) &         6223(258) &         2.785(73) &         0.261(22) \\
\hline
\multirow{1}{*}{$N_{>4\sigma}$}
& GBDT &                -- &                -- &                -- &                -- &                -- &                 1 \\
\hline\hline\\
\multicolumn{8}{c}{N$\approx$5000, D=5, $||\mathbf{c}||_1$=3.0}\\\hline\hline
\multicolumn{2}{c|}{Integrand family} & 1 & 2 & 3 & 4 & 5 & 6 \\\hline
\multicolumn{2}{c|}{$N$}&          5343(34) &       5201.9(7.8) &       5256.9(4.0) &       5209.7(6.5) &       5124.1(3.9) &          5759(11) \\
\multicolumn{2}{c|}{$N_{\textrm{train}}$}&          2653(13) &       2526.4(3.5) &       3716.3(3.4) &       2535.2(4.5) &       2493.4(1.8) &          1742(16) \\
\multicolumn{2}{c|}{$N_{\textrm{crxn}}$}&          2720(33) &       2751.8(9.0) &       1568.3(7.9) &       2756.3(3.4) &          2585(18) &          4053(31) \\
\hline
\multirow{4}{*}{$\sigma_I/|I|$}
& VEG & $1.1(4)\hspace{-0.3em}\times\hspace{-0.3em} 10^{-2}$ & $5.6(1)\hspace{-0.3em}\times\hspace{-0.3em} 10^{-4}$ & $2.04(2)\hspace{-0.3em}\times\hspace{-0.3em} 10^{-3}$ & $5.8(1)\hspace{-0.3em}\times\hspace{-0.3em} 10^{-4}$ & $6.6(1)\hspace{-0.3em}\times\hspace{-0.3em} 10^{-4}$ & $5.1(3)\hspace{-0.3em}\times\hspace{-0.3em} 10^{-3}$ \\
& MLP & $3(1)\hspace{-0.3em}\times\hspace{-0.3em} 10^{-4}$ & $6.6(2)\hspace{-0.3em}\times\hspace{-0.3em} 10^{-5}$ & $3.9(7)\hspace{-0.3em}\times\hspace{-0.3em} 10^{-4}$ & $7.5(2)\hspace{-0.3em}\times\hspace{-0.3em} 10^{-5}$ & $1.27(5)\hspace{-0.3em}\times\hspace{-0.3em} 10^{-4}$ & $5.1(5)\hspace{-0.3em}\times\hspace{-0.3em} 10^{-3}$ \\
& GBDT & $6(3)\hspace{-0.3em}\times\hspace{-0.3em} 10^{-3}$ & $1.33(3)\hspace{-0.3em}\times\hspace{-0.3em} 10^{-4}$ & $1.85(3)\hspace{-0.3em}\times\hspace{-0.3em} 10^{-3}$ & $1.50(3)\hspace{-0.3em}\times\hspace{-0.3em} 10^{-4}$ & $4.10(7)\hspace{-0.3em}\times\hspace{-0.3em} 10^{-4}$ & $8.0(2)\hspace{-0.3em}\times\hspace{-0.3em} 10^{-4}$ \\
& GP & $5(2)\hspace{-0.3em}\times\hspace{-0.3em} 10^{-5}$ & $4.8(3)\hspace{-0.3em}\times\hspace{-0.3em} 10^{-7}$ & $9.6(1)\hspace{-0.3em}\times\hspace{-0.3em} 10^{-6}$ & $5.4(3)\hspace{-0.3em}\times\hspace{-0.3em} 10^{-7}$ & $6.1(1)\hspace{-0.3em}\times\hspace{-0.3em} 10^{-4}$ & $2.6(3)\hspace{-0.3em}\times\hspace{-0.3em} 10^{-2}$ \\
\hline
\multirow{4}{*}{Gain}
& MLP &         32.7(1.1) &          8.73(28) &          6.59(31) &          7.87(21) &          5.50(19) &         1.141(66) \\
& GBDT &          2.18(11) &          4.33(16) &         1.129(36) &          3.95(14) &         1.647(45) &          6.49(42) \\
& GP &        196.8(1.7) &          1321(67) &        211.46(98) &          1151(47) &         1.105(30) &         0.253(19) \\
\hline
\multirow{1}{*}{$N_{>4\sigma}$}
& VEG &                -- &                -- &                -- &                -- &                -- &                 2 \\
\hline\hline\\
\multicolumn{8}{c}{N$\approx$5000, D=5, $||\mathbf{c}||_1$=8.0}\\\hline\hline
\multicolumn{2}{c|}{Integrand family} & 1 & 2 & 3 & 4 & 5 & 6 \\\hline
\multicolumn{2}{c|}{$N$}&       5245.0(6.9) &       5074.6(5.9) &       5415.2(5.5) &          5280(19) &       5113.2(8.7) &          5669(22) \\
\multicolumn{2}{c|}{$N_{\textrm{train}}$}&       2604.3(3.7) &       2513.1(4.7) &       3860.6(5.2) &       2567.0(8.1) &       2544.3(5.5) &          1801(15) \\
\multicolumn{2}{c|}{$N_{\textrm{crxn}}$}&          2609(20) &       2540.9(7.8) &       1535.8(7.9) &          2700(24) &          2582(11) &          3883(25) \\
\hline
\multirow{4}{*}{$\sigma_I/|I|$}
& VEG & $5(2)\hspace{-0.3em}\times\hspace{-0.3em} 10^{-2}$ & $9.1(1)\hspace{-0.3em}\times\hspace{-0.3em} 10^{-4}$ & $4.38(3)\hspace{-0.3em}\times\hspace{-0.3em} 10^{-3}$ & $1.74(6)\hspace{-0.3em}\times\hspace{-0.3em} 10^{-3}$ & $1.03(2)\hspace{-0.3em}\times\hspace{-0.3em} 10^{-3}$ & $6.2(3)\hspace{-0.3em}\times\hspace{-0.3em} 10^{-3}$ \\
& MLP & $1.3(7)\hspace{-0.3em}\times\hspace{-0.3em} 10^{-3}$ & $2.71(4)\hspace{-0.3em}\times\hspace{-0.3em} 10^{-4}$ & $8(1)\hspace{-0.3em}\times\hspace{-0.3em} 10^{-4}$ & $3.94(8)\hspace{-0.3em}\times\hspace{-0.3em} 10^{-4}$ & $3.5(1)\hspace{-0.3em}\times\hspace{-0.3em} 10^{-4}$ & $8.5(9)\hspace{-0.3em}\times\hspace{-0.3em} 10^{-3}$ \\
& GBDT & $9(9)\hspace{-0.3em}\times\hspace{-0.3em} 10^{-1}$ & $6.4(1)\hspace{-0.3em}\times\hspace{-0.3em} 10^{-4}$ & $4.5(1)\hspace{-0.3em}\times\hspace{-0.3em} 10^{-3}$ & $8.4(1)\hspace{-0.3em}\times\hspace{-0.3em} 10^{-4}$ & $1.17(2)\hspace{-0.3em}\times\hspace{-0.3em} 10^{-3}$ & $2.07(6)\hspace{-0.3em}\times\hspace{-0.3em} 10^{-3}$ \\
& GP & $2(1)\hspace{-0.3em}\times\hspace{-0.3em} 10^{-4}$ & $4.8(8)\hspace{-0.3em}\times\hspace{-0.3em} 10^{-5}$ & $3.3(1)\hspace{-0.3em}\times\hspace{-0.3em} 10^{-5}$ & $5.3(3)\hspace{-0.3em}\times\hspace{-0.3em} 10^{-6}$ & $1.98(4)\hspace{-0.3em}\times\hspace{-0.3em} 10^{-3}$ & $3.3(3)\hspace{-0.3em}\times\hspace{-0.3em} 10^{-2}$ \\
\hline
\multirow{4}{*}{Gain}
& MLP &         35.72(90) &         3.381(81) &          7.47(46) &          4.52(20) &          3.09(11) &         0.942(72) \\
& GBDT &         1.200(43) &         1.471(64) &         0.997(33) &          2.15(13) &         0.901(37) &          3.05(17) \\
& GP &        181.32(57) &         47.2(7.9) &        137.6(4.5) &        346.2(9.4) &         0.526(11) &         0.220(13) \\
\hline
\multirow{1}{*}{$N_{>4\sigma}$}
& VEG &                -- &                -- &                -- &                -- &                -- &                 2 \\
\hline\hline    
    \end{tabular}
    \caption{Number of integrand evaluations ($N$, $N_{\textrm{train}}$, and $N_{\textrm{crxn}}$), precision of the integral ($\sigma_I/|I|$), gain defined in Eq.~\protect\eqref{eq:gain}, and $N_{>4\sigma}$ of VEGAS (VEG) and ML (MLP, GBDT, and GP) integration algorithms for $N\approx 5000$ and $D=5$. The results are averaged over 36 random samples. The numbers in the parentheses are the standard deviation of the mean. Here, $N_{>4\sigma}$ is the number of integration results that are more than 4$\sigma$ away from the true answer out of the 36 samples; only the non-zero values are presented.}
    \label{tab:res5000-5}
\end{table}
%%%%%%%%%%%%%%%%%%%%%%%%%%%%%%%%%%%%%%%%%%
\begin{table}[htbp]
\footnotesize
    \centering
    \begin{tabular}{cc|cccccc}
\multicolumn{8}{c}{N$\approx$5000, D=8, $||\mathbf{c}||_1$=1.0}\\\hline\hline
\multicolumn{2}{c|}{Integrand family} & 1 & 2 & 3 & 4 & 5 & 6 \\\hline
\multicolumn{2}{c|}{$N$}&           5000(0) &           5000(0) &           5000(0) &           5000(0) &           5000(0) &           5000(0) \\
\multicolumn{2}{c|}{$N_{\textrm{train}}$}&           2500(0) &           2500(0) &           3500(0) &           2500(0) &           2500(0) &           1500(0) \\
\multicolumn{2}{c|}{$N_{\textrm{crxn}}$}&           2500(0) &           2500(0) &           1500(0) &           2500(0) &           2500(0) &           3500(0) \\
\hline
\multirow{4}{*}{$\sigma_I/|I|$}
& VEG & $9(6)\hspace{-0.3em}\times\hspace{-0.3em} 10^{-3}$ & $5.3(1)\hspace{-0.3em}\times\hspace{-0.3em} 10^{-4}$ & $1.42(1)\hspace{-0.3em}\times\hspace{-0.3em} 10^{-3}$ & $5.21(9)\hspace{-0.3em}\times\hspace{-0.3em} 10^{-4}$ & $7.79(8)\hspace{-0.3em}\times\hspace{-0.3em} 10^{-4}$ & $5.5(4)\hspace{-0.3em}\times\hspace{-0.3em} 10^{-3}$ \\
& MLP & $2(1)\hspace{-0.3em}\times\hspace{-0.3em} 10^{-4}$ & $1.17(4)\hspace{-0.3em}\times\hspace{-0.3em} 10^{-5}$ & $2.47(7)\hspace{-0.3em}\times\hspace{-0.3em} 10^{-4}$ & $1.09(2)\hspace{-0.3em}\times\hspace{-0.3em} 10^{-5}$ & $9.0(2)\hspace{-0.3em}\times\hspace{-0.3em} 10^{-5}$ & $9(1)\hspace{-0.3em}\times\hspace{-0.3em} 10^{-3}$ \\
& GBDT & $4(2)\hspace{-0.3em}\times\hspace{-0.3em} 10^{-3}$ & $1.78(2)\hspace{-0.3em}\times\hspace{-0.3em} 10^{-5}$ & $1.90(2)\hspace{-0.3em}\times\hspace{-0.3em} 10^{-3}$ & $1.78(2)\hspace{-0.3em}\times\hspace{-0.3em} 10^{-5}$ & $2.03(2)\hspace{-0.3em}\times\hspace{-0.3em} 10^{-4}$ & $6.4(2)\hspace{-0.3em}\times\hspace{-0.3em} 10^{-4}$ \\
& GP & $5(3)\hspace{-0.3em}\times\hspace{-0.3em} 10^{-5}$ & $6.9(4)\hspace{-0.3em}\times\hspace{-0.3em} 10^{-8}$ & $6.9(1)\hspace{-0.3em}\times\hspace{-0.3em} 10^{-6}$ & $5.9(2)\hspace{-0.3em}\times\hspace{-0.3em} 10^{-8}$ & $2.45(5)\hspace{-0.3em}\times\hspace{-0.3em} 10^{-4}$ & $3.0(4)\hspace{-0.3em}\times\hspace{-0.3em} 10^{-2}$ \\
\hline
\multirow{4}{*}{Gain}
& MLP &         40.5(4.0) &         46.37(86) &          5.89(16) &         48.03(96) &          8.88(29) &         0.714(36) \\
& GBDT &          4.53(37) &         30.15(95) &         0.752(17) &         29.19(52) &         3.863(76) &          8.62(59) \\
& GP &           352(28) &         8292(262) &        206.8(2.5) &         8918(177) &          3.25(10) &         0.234(16) \\
\hline
\multirow{2}{*}{$N_{>4\sigma}$}
& VEG &                -- &                -- &                -- &                -- &                -- &                 1 \\
& GBDT &                -- &                -- &                -- &                -- &                -- &                 1 \\
\hline\hline\\
\multicolumn{8}{c}{N$\approx$5000, D=8, $||\mathbf{c}||_1$=3.0}\\\hline\hline
\multicolumn{2}{c|}{Integrand family} & 1 & 2 & 3 & 4 & 5 & 6 \\\hline
\multicolumn{2}{c|}{$N$}&           5000(0) &           5000(0) &           5000(0) &           5000(0) &           5000(0) &           5000(0) \\
\multicolumn{2}{c|}{$N_{\textrm{train}}$}&           2500(0) &           2500(0) &           3500(0) &           2500(0) &           2500(0) &           1500(0) \\
\multicolumn{2}{c|}{$N_{\textrm{crxn}}$}&           2500(0) &           2500(0) &           1500(0) &           2500(0) &           2500(0) &           3500(0) \\
\hline
\multirow{4}{*}{$\sigma_I/|I|$}
& VEG & $1.5(6)\hspace{-0.3em}\times\hspace{-0.3em} 10^{-2}$ & $8.1(1)\hspace{-0.3em}\times\hspace{-0.3em} 10^{-4}$ & $3.47(4)\hspace{-0.3em}\times\hspace{-0.3em} 10^{-3}$ & $8.10(9)\hspace{-0.3em}\times\hspace{-0.3em} 10^{-4}$ & $9.4(1)\hspace{-0.3em}\times\hspace{-0.3em} 10^{-4}$ & $5.8(4)\hspace{-0.3em}\times\hspace{-0.3em} 10^{-3}$ \\
& MLP & $3(1)\hspace{-0.3em}\times\hspace{-0.3em} 10^{-4}$ & $9.1(2)\hspace{-0.3em}\times\hspace{-0.3em} 10^{-5}$ & $6.0(1)\hspace{-0.3em}\times\hspace{-0.3em} 10^{-4}$ & $9.2(2)\hspace{-0.3em}\times\hspace{-0.3em} 10^{-5}$ & $2.71(9)\hspace{-0.3em}\times\hspace{-0.3em} 10^{-4}$ & $1.1(1)\hspace{-0.3em}\times\hspace{-0.3em} 10^{-2}$ \\
& GBDT & $1.2(6)\hspace{-0.3em}\times\hspace{-0.3em} 10^{-2}$ & $1.49(2)\hspace{-0.3em}\times\hspace{-0.3em} 10^{-4}$ & $5.95(8)\hspace{-0.3em}\times\hspace{-0.3em} 10^{-3}$ & $1.58(2)\hspace{-0.3em}\times\hspace{-0.3em} 10^{-4}$ & $6.17(7)\hspace{-0.3em}\times\hspace{-0.3em} 10^{-4}$ & $1.33(4)\hspace{-0.3em}\times\hspace{-0.3em} 10^{-3}$ \\
& GP & $1.0(4)\hspace{-0.3em}\times\hspace{-0.3em} 10^{-4}$ & $1.1(2)\hspace{-0.3em}\times\hspace{-0.3em} 10^{-6}$ & $2.67(5)\hspace{-0.3em}\times\hspace{-0.3em} 10^{-5}$ & $5.2(1)\hspace{-0.3em}\times\hspace{-0.3em} 10^{-7}$ & $7.6(1)\hspace{-0.3em}\times\hspace{-0.3em} 10^{-4}$ & $3.1(4)\hspace{-0.3em}\times\hspace{-0.3em} 10^{-2}$ \\
\hline
\multirow{4}{*}{Gain}
& MLP &         30.2(2.1) &          9.17(22) &          5.83(14) &          8.88(19) &          3.64(14) &         0.603(29) \\
& GBDT &         1.970(74) &          5.52(13) &         0.592(15) &         5.143(86) &         1.542(27) &          4.32(22) \\
& GP &        140.5(1.6) &         1219(100) &        130.4(1.2) &          1606(46) &         1.270(41) &         0.231(16) \\
\hline
\multirow{1}{*}{$N_{>4\sigma}$}
& VEG &                -- &                -- &                -- &                -- &                -- &                 2 \\
\hline\hline\\
\multicolumn{8}{c}{N$\approx$5000, D=8, $||\mathbf{c}||_1$=8.0}\\\hline\hline
\multicolumn{2}{c|}{Integrand family} & 1 & 2 & 3 & 4 & 5 & 6 \\\hline
\multicolumn{2}{c|}{$N$}&           5000(0) &           5000(0) &           5000(0) &           5000(0) &           5000(0) &           5000(0) \\
\multicolumn{2}{c|}{$N_{\textrm{train}}$}&           2500(0) &           2500(0) &           3500(0) &           2500(0) &           2500(0) &           1500(0) \\
\multicolumn{2}{c|}{$N_{\textrm{crxn}}$}&           2500(0) &           2500(0) &           1500(0) &           2500(0) &           2500(0) &           3500(0) \\
\hline
\multirow{4}{*}{$\sigma_I/|I|$}
& VEG & $3(1)\hspace{-0.3em}\times\hspace{-0.3em} 10^{-2}$ & $1.12(1)\hspace{-0.3em}\times\hspace{-0.3em} 10^{-3}$ & $8.1(1)\hspace{-0.3em}\times\hspace{-0.3em} 10^{-3}$ & $1.34(2)\hspace{-0.3em}\times\hspace{-0.3em} 10^{-3}$ & $1.24(1)\hspace{-0.3em}\times\hspace{-0.3em} 10^{-3}$ & $8.0(3)\hspace{-0.3em}\times\hspace{-0.3em} 10^{-3}$ \\
& MLP & $1(1)\hspace{-0.3em}\times\hspace{-0.3em} 10^{-3}$ & $4.38(6)\hspace{-0.3em}\times\hspace{-0.3em} 10^{-4}$ & $1.24(5)\hspace{-0.3em}\times\hspace{-0.3em} 10^{-3}$ & $5.85(9)\hspace{-0.3em}\times\hspace{-0.3em} 10^{-4}$ & $8.0(2)\hspace{-0.3em}\times\hspace{-0.3em} 10^{-4}$ & $1.3(1)\hspace{-0.3em}\times\hspace{-0.3em} 10^{-2}$ \\
& GBDT & $1(1)\hspace{-0.3em}\times\hspace{-0.3em} 10^{-1}$ & $8.0(1)\hspace{-0.3em}\times\hspace{-0.3em} 10^{-4}$ & $1.85(4)\hspace{-0.3em}\times\hspace{-0.3em} 10^{-2}$ & $1.02(1)\hspace{-0.3em}\times\hspace{-0.3em} 10^{-3}$ & $1.77(2)\hspace{-0.3em}\times\hspace{-0.3em} 10^{-3}$ & $3.47(9)\hspace{-0.3em}\times\hspace{-0.3em} 10^{-3}$ \\
& GP & $5(2)\hspace{-0.3em}\times\hspace{-0.3em} 10^{-4}$ & $1.0(1)\hspace{-0.3em}\times\hspace{-0.3em} 10^{-4}$ & $1.55(9)\hspace{-0.3em}\times\hspace{-0.3em} 10^{-4}$ & $4.3(1)\hspace{-0.3em}\times\hspace{-0.3em} 10^{-6}$ & $2.23(4)\hspace{-0.3em}\times\hspace{-0.3em} 10^{-3}$ & $3.9(4)\hspace{-0.3em}\times\hspace{-0.3em} 10^{-2}$ \\
\hline
\multirow{4}{*}{Gain}
& MLP &         32.98(82) &         2.588(49) &          7.03(31) &         2.309(45) &         1.613(63) &         0.741(51) \\
& GBDT &         1.191(27) &         1.420(44) &         0.450(10) &         1.319(33) &         0.707(13) &         2.276(71) \\
& GP &        106.5(2.8) &         20.2(2.3) &         57.2(2.5) &        319.4(6.5) &         0.568(16) &         0.252(18) \\
\hline
\multirow{2}{*}{$N_{>4\sigma}$}
& VEG &                -- &                -- &                -- &                -- &                -- &                 2 \\
& MLP &                -- &                -- &                 1 &                -- &                -- &                -- \\
\hline\hline
    \end{tabular}
    \caption{Number of integrand evaluations, precision of the integral, precision gain, and $N_{>4\sigma}$ for $N\approx 5000$ and $D=8$. The notations are the same as Table~\ref{tab:res5000-5}.}
    \label{tab:res5000-8}
\end{table}
%%%%%%%%%%%%%%%%%%%%%%%%%%%%%%%%%%%%%%%%%%
\begin{table}[htbp]
\footnotesize
    \centering
    \begin{tabular}{cc|cccccc}
\multicolumn{8}{c}{N$\approx$5000, D=10, $||\mathbf{c}||_1$=1.0}\\\hline\hline
\multicolumn{2}{c|}{Integrand family} & 1 & 2 & 3 & 4 & 5 & 6 \\\hline
\multicolumn{2}{c|}{$N$}&           5000(0) &           5000(0) &           5000(0) &           5000(0) &           5000(0) &           5000(0) \\
\multicolumn{2}{c|}{$N_{\textrm{train}}$}&           2500(0) &           2500(0) &           3500(0) &           2500(0) &           2500(0) &           1500(0) \\
\multicolumn{2}{c|}{$N_{\textrm{crxn}}$}&           2500(0) &           2500(0) &           1500(0) &           2500(0) &           2500(0) &           3500(0) \\
\hline
\multirow{4}{*}{$\sigma_I/|I|$}
& VEG & $6(4)\hspace{-0.3em}\times\hspace{-0.3em} 10^{-3}$ & $4.8(1)\hspace{-0.3em}\times\hspace{-0.3em} 10^{-4}$ & $1.60(1)\hspace{-0.3em}\times\hspace{-0.3em} 10^{-3}$ & $4.48(9)\hspace{-0.3em}\times\hspace{-0.3em} 10^{-4}$ & $8.85(8)\hspace{-0.3em}\times\hspace{-0.3em} 10^{-4}$ & $6.2(4)\hspace{-0.3em}\times\hspace{-0.3em} 10^{-3}$ \\
& MLP & $2(1)\hspace{-0.3em}\times\hspace{-0.3em} 10^{-4}$ & $1.16(4)\hspace{-0.3em}\times\hspace{-0.3em} 10^{-5}$ & $3.2(1)\hspace{-0.3em}\times\hspace{-0.3em} 10^{-4}$ & $1.08(2)\hspace{-0.3em}\times\hspace{-0.3em} 10^{-5}$ & $1.42(4)\hspace{-0.3em}\times\hspace{-0.3em} 10^{-4}$ & $1.2(1)\hspace{-0.3em}\times\hspace{-0.3em} 10^{-2}$ \\
& GBDT & $4(3)\hspace{-0.3em}\times\hspace{-0.3em} 10^{-3}$ & $1.68(2)\hspace{-0.3em}\times\hspace{-0.3em} 10^{-5}$ & $2.83(3)\hspace{-0.3em}\times\hspace{-0.3em} 10^{-3}$ & $1.66(2)\hspace{-0.3em}\times\hspace{-0.3em} 10^{-5}$ & $2.32(2)\hspace{-0.3em}\times\hspace{-0.3em} 10^{-4}$ & $7.6(4)\hspace{-0.3em}\times\hspace{-0.3em} 10^{-4}$ \\
& GP & $6(4)\hspace{-0.3em}\times\hspace{-0.3em} 10^{-5}$ & $5.9(4)\hspace{-0.3em}\times\hspace{-0.3em} 10^{-8}$ & $9.0(1)\hspace{-0.3em}\times\hspace{-0.3em} 10^{-6}$ & $5.0(1)\hspace{-0.3em}\times\hspace{-0.3em} 10^{-8}$ & $2.54(5)\hspace{-0.3em}\times\hspace{-0.3em} 10^{-4}$ & $3.1(4)\hspace{-0.3em}\times\hspace{-0.3em} 10^{-2}$ \\
\hline
\multirow{4}{*}{Gain}
& MLP &         35.3(3.1) &         41.41(88) &          5.05(15) &         41.78(99) &          6.48(23) &         0.593(31) \\
& GBDT &          4.37(38) &         28.57(93) &         0.572(12) &         26.95(58) &         3.825(66) &          8.21(42) \\
& GP &           345(29) &         8533(250) &        178.4(2.2) &         8941(189) &          3.55(11) &         0.245(15) \\
\hline
\multirow{1}{*}{$N_{>4\sigma}$}
& VEG &                -- &                -- &                -- &                -- &                -- &                 1 \\
\hline\hline\\
\multicolumn{8}{c}{N$\approx$5000, D=10, $||\mathbf{c}||_1$=3.0}\\\hline\hline
\multicolumn{2}{c|}{Integrand family} & 1 & 2 & 3 & 4 & 5 & 6 \\\hline
\multicolumn{2}{c|}{$N$}&           5000(0) &           5000(0) &           5000(0) &           5000(0) &           5000(0) &           5000(0) \\
\multicolumn{2}{c|}{$N_{\textrm{train}}$}&           2500(0) &           2500(0) &           3500(0) &           2500(0) &           2500(0) &           1500(0) \\
\multicolumn{2}{c|}{$N_{\textrm{crxn}}$}&           2500(0) &           2500(0) &           1500(0) &           2500(0) &           2500(0) &           3500(0) \\
\hline
\multirow{4}{*}{$\sigma_I/|I|$}
& VEG & $1.3(5)\hspace{-0.3em}\times\hspace{-0.3em} 10^{-2}$ & $8.91(9)\hspace{-0.3em}\times\hspace{-0.3em} 10^{-4}$ & $3.94(5)\hspace{-0.3em}\times\hspace{-0.3em} 10^{-3}$ & $8.83(8)\hspace{-0.3em}\times\hspace{-0.3em} 10^{-4}$ & $1.06(1)\hspace{-0.3em}\times\hspace{-0.3em} 10^{-3}$ & $6.7(5)\hspace{-0.3em}\times\hspace{-0.3em} 10^{-3}$ \\
& MLP & $3(1)\hspace{-0.3em}\times\hspace{-0.3em} 10^{-4}$ & $9.6(3)\hspace{-0.3em}\times\hspace{-0.3em} 10^{-5}$ & $8.4(2)\hspace{-0.3em}\times\hspace{-0.3em} 10^{-4}$ & $9.5(1)\hspace{-0.3em}\times\hspace{-0.3em} 10^{-5}$ & $4.4(1)\hspace{-0.3em}\times\hspace{-0.3em} 10^{-4}$ & $1.3(1)\hspace{-0.3em}\times\hspace{-0.3em} 10^{-2}$ \\
& GBDT & $1.0(4)\hspace{-0.3em}\times\hspace{-0.3em} 10^{-2}$ & $1.42(2)\hspace{-0.3em}\times\hspace{-0.3em} 10^{-4}$ & $9.3(1)\hspace{-0.3em}\times\hspace{-0.3em} 10^{-3}$ & $1.47(1)\hspace{-0.3em}\times\hspace{-0.3em} 10^{-4}$ & $7.11(7)\hspace{-0.3em}\times\hspace{-0.3em} 10^{-4}$ & $1.49(6)\hspace{-0.3em}\times\hspace{-0.3em} 10^{-3}$ \\
& GP & $1.1(5)\hspace{-0.3em}\times\hspace{-0.3em} 10^{-4}$ & $2.0(5)\hspace{-0.3em}\times\hspace{-0.3em} 10^{-6}$ & $4.0(1)\hspace{-0.3em}\times\hspace{-0.3em} 10^{-5}$ & $4.4(1)\hspace{-0.3em}\times\hspace{-0.3em} 10^{-7}$ & $7.7(1)\hspace{-0.3em}\times\hspace{-0.3em} 10^{-4}$ & $3.3(4)\hspace{-0.3em}\times\hspace{-0.3em} 10^{-2}$ \\
\hline
\multirow{4}{*}{Gain}
& MLP &         25.2(1.8) &          9.51(27) &          4.81(15) &          9.39(20) &         2.477(75) &         0.573(32) \\
& GBDT &         1.759(57) &          6.31(12) &        0.4266(88) &         6.002(87) &         1.511(28) &          4.34(23) \\
& GP &        118.7(1.3) &          972(106) &         98.7(1.4) &          2028(56) &         1.405(40) &         0.247(15) \\
\hline
\multirow{1}{*}{$N_{>4\sigma}$}
& VEG &                -- &                -- &                -- &                -- &                -- &                 3 \\
\hline\hline\\
\multicolumn{8}{c}{N$\approx$5000, D=10, $||\mathbf{c}||_1$=8.0}\\\hline\hline
\multicolumn{2}{c|}{Integrand family} & 1 & 2 & 3 & 4 & 5 & 6 \\\hline
\multicolumn{2}{c|}{$N$}&           5000(0) &           5000(0) &           5000(0) &           5000(0) &           5000(0) &           5000(0) \\
\multicolumn{2}{c|}{$N_{\textrm{train}}$}&           2500(0) &           2500(0) &           3500(0) &           2500(0) &           2500(0) &           1500(0) \\
\multicolumn{2}{c|}{$N_{\textrm{crxn}}$}&           2500(0) &           2500(0) &           1500(0) &           2500(0) &           2500(0) &           3500(0) \\
\hline
\multirow{4}{*}{$\sigma_I/|I|$}
& VEG & $3(1)\hspace{-0.3em}\times\hspace{-0.3em} 10^{-2}$ & $1.20(1)\hspace{-0.3em}\times\hspace{-0.3em} 10^{-3}$ & $9.3(1)\hspace{-0.3em}\times\hspace{-0.3em} 10^{-3}$ & $1.30(1)\hspace{-0.3em}\times\hspace{-0.3em} 10^{-3}$ & $1.36(1)\hspace{-0.3em}\times\hspace{-0.3em} 10^{-3}$ & $8.1(4)\hspace{-0.3em}\times\hspace{-0.3em} 10^{-3}$ \\
& MLP & $1.6(8)\hspace{-0.3em}\times\hspace{-0.3em} 10^{-3}$ & $5.05(7)\hspace{-0.3em}\times\hspace{-0.3em} 10^{-4}$ & $1.70(8)\hspace{-0.3em}\times\hspace{-0.3em} 10^{-3}$ & $6.2(1)\hspace{-0.3em}\times\hspace{-0.3em} 10^{-4}$ & $1.20(4)\hspace{-0.3em}\times\hspace{-0.3em} 10^{-3}$ & $1.7(2)\hspace{-0.3em}\times\hspace{-0.3em} 10^{-2}$ \\
& GBDT & $6(4)\hspace{-0.3em}\times\hspace{-0.3em} 10^{-2}$ & $8.2(1)\hspace{-0.3em}\times\hspace{-0.3em} 10^{-4}$ & $3.33(7)\hspace{-0.3em}\times\hspace{-0.3em} 10^{-2}$ & $9.9(1)\hspace{-0.3em}\times\hspace{-0.3em} 10^{-4}$ & $2.03(2)\hspace{-0.3em}\times\hspace{-0.3em} 10^{-3}$ & $3.9(1)\hspace{-0.3em}\times\hspace{-0.3em} 10^{-3}$ \\
& GP & $8(4)\hspace{-0.3em}\times\hspace{-0.3em} 10^{-4}$ & $1.1(1)\hspace{-0.3em}\times\hspace{-0.3em} 10^{-4}$ & $2.8(1)\hspace{-0.3em}\times\hspace{-0.3em} 10^{-4}$ & $3.45(9)\hspace{-0.3em}\times\hspace{-0.3em} 10^{-6}$ & $2.23(4)\hspace{-0.3em}\times\hspace{-0.3em} 10^{-3}$ & $4.2(5)\hspace{-0.3em}\times\hspace{-0.3em} 10^{-2}$ \\
\hline
\multirow{4}{*}{Gain}
& MLP &         25.99(72) &         2.389(34) &          5.87(26) &         2.102(38) &         1.184(50) &         0.595(39) \\
& GBDT &         1.076(20) &         1.472(40) &        0.2780(81) &         1.321(20) &         0.676(14) &         2.005(72) \\
& GP &         53.6(1.5) &         17.4(1.7) &         34.6(1.4) &        385.8(9.8) &         0.622(16) &         0.241(17) \\
\hline
\multirow{3}{*}{$N_{>4\sigma}$}
& VEG &                -- &                -- &                -- &                -- &                -- &                 3 \\
& MLP &                -- &                -- &                 5 &                -- &                -- &                -- \\
& GBDT &                -- &                -- &                 1 &                -- &                -- &                -- \\
\hline\hline
    \end{tabular}
    \caption{Number of integrand evaluations, precision of the integral, precision gain, and $N_{>4\sigma}$ for $N\approx 5000$ and $D=10$. The notations are the same as Table~\ref{tab:res5000-5}.}
    \label{tab:res5000-10}
\end{table}
%%%%%%%%%%%%%%%%%%%%%%%%%%%%%%%%%%%%%%%%%%
\begin{table}[htbp]
\footnotesize
    \centering
    \begin{tabular}{cc|cccccc}
\multicolumn{8}{c}{N$\approx$10000, D=5, $||\mathbf{c}||_1$=1.0}\\\hline\hline
\multicolumn{2}{c|}{Integrand family} & 1 & 2 & 3 & 4 & 5 & 6 \\\hline
\multicolumn{2}{c|}{$N$}&         11720(59) &         12372(30) &      11269.9(5.5) &         12387(22) &         11712(14) &         13173(95) \\
\multicolumn{2}{c|}{$N_{\textrm{train}}$}&          5746(33) &          6060(19) &       7851.1(4.4) &          6066(18) &       5594.3(6.8) &          4083(13) \\
\multicolumn{2}{c|}{$N_{\textrm{crxn}}$}&          6052(19) &          6200(11) &          3444(28) &       6201.4(8.9) &       6164.3(9.1) &         9181(103) \\
\hline
\multirow{4}{*}{$\sigma_I/|I|$}
& VEG & $3(2)\hspace{-0.3em}\times\hspace{-0.3em} 10^{-3}$ & $1.76(3)\hspace{-0.3em}\times\hspace{-0.3em} 10^{-4}$ & $4.47(5)\hspace{-0.3em}\times\hspace{-0.3em} 10^{-4}$ & $1.72(4)\hspace{-0.3em}\times\hspace{-0.3em} 10^{-4}$ & $2.43(3)\hspace{-0.3em}\times\hspace{-0.3em} 10^{-4}$ & $2.7(1)\hspace{-0.3em}\times\hspace{-0.3em} 10^{-3}$ \\
& MLP & $7(4)\hspace{-0.3em}\times\hspace{-0.3em} 10^{-5}$ & $4.6(1)\hspace{-0.3em}\times\hspace{-0.3em} 10^{-6}$ & $5.9(1)\hspace{-0.3em}\times\hspace{-0.3em} 10^{-5}$ & $4.4(1)\hspace{-0.3em}\times\hspace{-0.3em} 10^{-6}$ & $1.51(5)\hspace{-0.3em}\times\hspace{-0.3em} 10^{-5}$ & $2.0(2)\hspace{-0.3em}\times\hspace{-0.3em} 10^{-3}$ \\
& GBDT & $1.5(9)\hspace{-0.3em}\times\hspace{-0.3em} 10^{-3}$ & $9.3(1)\hspace{-0.3em}\times\hspace{-0.3em} 10^{-6}$ & $4.20(5)\hspace{-0.3em}\times\hspace{-0.3em} 10^{-4}$ & $9.5(2)\hspace{-0.3em}\times\hspace{-0.3em} 10^{-6}$ & $7.5(1)\hspace{-0.3em}\times\hspace{-0.3em} 10^{-5}$ & $1.91(9)\hspace{-0.3em}\times\hspace{-0.3em} 10^{-4}$ \\
& GP & $1(1)\hspace{-0.3em}\times\hspace{-0.3em} 10^{-5}$ & $3.4(2)\hspace{-0.3em}\times\hspace{-0.3em} 10^{-8}$ & $1.50(2)\hspace{-0.3em}\times\hspace{-0.3em} 10^{-6}$ & $3.3(1)\hspace{-0.3em}\times\hspace{-0.3em} 10^{-8}$ & $9.7(2)\hspace{-0.3em}\times\hspace{-0.3em} 10^{-5}$ & $1.5(2)\hspace{-0.3em}\times\hspace{-0.3em} 10^{-2}$ \\
\hline
\multirow{4}{*}{Gain}
& MLP &         41.7(3.9) &         39.9(1.3) &          7.65(16) &         39.2(1.2) &         16.67(54) &         1.531(75) \\
& GBDT &          4.51(36) &         19.22(54) &         1.076(27) &         18.42(62) &         3.288(97) &         14.52(66) \\
& GP &           357(24) &         5770(311) &        298.9(2.9) &         5497(235) &         2.546(65) &         0.243(18) \\
\hline
\multirow{1}{*}{$N_{>4\sigma}$}
& -- &                -- &                -- &                -- &                -- &                -- &                -- \\
\hline\hline\\
\multicolumn{8}{c}{N$\approx$10000, D=5, $||\mathbf{c}||_1$=3.0}\\\hline\hline
\multicolumn{2}{c|}{Integrand family} & 1 & 2 & 3 & 4 & 5 & 6 \\\hline
\multicolumn{2}{c|}{$N$}&         11631(59) &         11511(21) &      11498.7(8.0) &         11483(18) &      11325.9(9.1) &         12751(39) \\
\multicolumn{2}{c|}{$N_{\textrm{train}}$}&          5743(20) &       5533.6(5.5) &       8110.2(6.6) &       5543.6(6.0) &       5493.7(3.5) &          3952(25) \\
\multicolumn{2}{c|}{$N_{\textrm{crxn}}$}&          5997(21) &          6008(10) &          3335(14) &       5986.5(9.4) &          5998(23) &          8842(60) \\
\hline
\multirow{4}{*}{$\sigma_I/|I|$}
& VEG & $5(2)\hspace{-0.3em}\times\hspace{-0.3em} 10^{-3}$ & $2.83(5)\hspace{-0.3em}\times\hspace{-0.3em} 10^{-4}$ & $1.07(1)\hspace{-0.3em}\times\hspace{-0.3em} 10^{-3}$ & $3.04(6)\hspace{-0.3em}\times\hspace{-0.3em} 10^{-4}$ & $3.40(5)\hspace{-0.3em}\times\hspace{-0.3em} 10^{-4}$ & $2.9(1)\hspace{-0.3em}\times\hspace{-0.3em} 10^{-3}$ \\
& MLP & $1.3(5)\hspace{-0.3em}\times\hspace{-0.3em} 10^{-4}$ & $3.29(6)\hspace{-0.3em}\times\hspace{-0.3em} 10^{-5}$ & $1.8(2)\hspace{-0.3em}\times\hspace{-0.3em} 10^{-4}$ & $3.9(1)\hspace{-0.3em}\times\hspace{-0.3em} 10^{-5}$ & $4.4(1)\hspace{-0.3em}\times\hspace{-0.3em} 10^{-5}$ & $2.6(3)\hspace{-0.3em}\times\hspace{-0.3em} 10^{-3}$ \\
& GBDT & $4(2)\hspace{-0.3em}\times\hspace{-0.3em} 10^{-3}$ & $7.5(1)\hspace{-0.3em}\times\hspace{-0.3em} 10^{-5}$ & $1.12(2)\hspace{-0.3em}\times\hspace{-0.3em} 10^{-3}$ & $8.3(1)\hspace{-0.3em}\times\hspace{-0.3em} 10^{-5}$ & $2.32(4)\hspace{-0.3em}\times\hspace{-0.3em} 10^{-4}$ & $3.9(1)\hspace{-0.3em}\times\hspace{-0.3em} 10^{-4}$ \\
& GP & $3(1)\hspace{-0.3em}\times\hspace{-0.3em} 10^{-5}$ & $2.4(1)\hspace{-0.3em}\times\hspace{-0.3em} 10^{-7}$ & $5.39(7)\hspace{-0.3em}\times\hspace{-0.3em} 10^{-6}$ & $3.0(1)\hspace{-0.3em}\times\hspace{-0.3em} 10^{-7}$ & $3.19(6)\hspace{-0.3em}\times\hspace{-0.3em} 10^{-4}$ & $1.6(2)\hspace{-0.3em}\times\hspace{-0.3em} 10^{-2}$ \\
\hline
\multirow{4}{*}{Gain}
& MLP &         35.43(98) &          8.67(18) &          6.90(28) &          7.76(17) &          7.88(25) &         1.262(68) \\
& GBDT &          2.05(11) &          3.84(14) &         0.975(27) &          3.70(12) &         1.503(54) &          7.47(34) \\
& GP &        174.7(1.6) &          1241(44) &        200.16(92) &          1084(44) &         1.076(25) &         0.234(15) \\
\hline
\multirow{1}{*}{$N_{>4\sigma}$}
& -- &                -- &                -- &                -- &                -- &                -- &                -- \\
\hline\hline\\
\multicolumn{8}{c}{N$\approx$10000, D=5, $||\mathbf{c}||_1$=8.0}\\\hline\hline
\multicolumn{2}{c|}{Integrand family} & 1 & 2 & 3 & 4 & 5 & 6 \\\hline
\multicolumn{2}{c|}{$N$}&         11420(11) &      11193.7(8.0) &         11814(13) &         11517(32) &         11244(14) &         12519(37) \\
\multicolumn{2}{c|}{$N_{\textrm{train}}$}&       5677.7(5.1) &       5531.6(5.8) &          8416(11) &          5645(15) &       5599.8(8.0) &          4008(32) \\
\multicolumn{2}{c|}{$N_{\textrm{crxn}}$}&          5831(37) &          5742(43) &          3456(18) &          5996(35) &          5517(22) &          8535(53) \\
\hline
\multirow{4}{*}{$\sigma_I/|I|$}
& VEG & $2.0(8)\hspace{-0.3em}\times\hspace{-0.3em} 10^{-2}$ & $4.64(7)\hspace{-0.3em}\times\hspace{-0.3em} 10^{-4}$ & $2.35(1)\hspace{-0.3em}\times\hspace{-0.3em} 10^{-3}$ & $8.5(3)\hspace{-0.3em}\times\hspace{-0.3em} 10^{-4}$ & $5.0(1)\hspace{-0.3em}\times\hspace{-0.3em} 10^{-4}$ & $3.7(1)\hspace{-0.3em}\times\hspace{-0.3em} 10^{-3}$ \\
& MLP & $5(2)\hspace{-0.3em}\times\hspace{-0.3em} 10^{-4}$ & $1.31(2)\hspace{-0.3em}\times\hspace{-0.3em} 10^{-4}$ & $4(1)\hspace{-0.3em}\times\hspace{-0.3em} 10^{-4}$ & $2.13(6)\hspace{-0.3em}\times\hspace{-0.3em} 10^{-4}$ & $1.32(4)\hspace{-0.3em}\times\hspace{-0.3em} 10^{-4}$ & $3.6(5)\hspace{-0.3em}\times\hspace{-0.3em} 10^{-3}$ \\
& GBDT & $2(1)\hspace{-0.3em}\times\hspace{-0.3em} 10^{-2}$ & $3.49(9)\hspace{-0.3em}\times\hspace{-0.3em} 10^{-4}$ & $2.91(6)\hspace{-0.3em}\times\hspace{-0.3em} 10^{-3}$ & $4.7(1)\hspace{-0.3em}\times\hspace{-0.3em} 10^{-4}$ & $6.8(1)\hspace{-0.3em}\times\hspace{-0.3em} 10^{-4}$ & $1.06(2)\hspace{-0.3em}\times\hspace{-0.3em} 10^{-3}$ \\
& GP & $1.4(6)\hspace{-0.3em}\times\hspace{-0.3em} 10^{-4}$ & $1.2(2)\hspace{-0.3em}\times\hspace{-0.3em} 10^{-5}$ & $2.0(3)\hspace{-0.3em}\times\hspace{-0.3em} 10^{-5}$ & $3.2(2)\hspace{-0.3em}\times\hspace{-0.3em} 10^{-6}$ & $1.08(2)\hspace{-0.3em}\times\hspace{-0.3em} 10^{-3}$ & $2.3(2)\hspace{-0.3em}\times\hspace{-0.3em} 10^{-2}$ \\
\hline
\multirow{4}{*}{Gain}
& MLP &         40.6(1.0) &         3.564(87) &          7.47(41) &          4.13(22) &          3.96(12) &         1.350(80) \\
& GBDT &         0.997(35) &         1.379(57) &         0.825(24) &          1.86(11) &         0.770(32) &          3.55(16) \\
& GP &        157.38(47) &           121(20) &        144.3(6.4) &        279.4(7.4) &        0.4730(90) &         0.205(14) \\
\hline
\multirow{1}{*}{$N_{>4\sigma}$}
& VEG &                -- &                -- &                -- &                -- &                -- &                 1 \\
\hline\hline
    \end{tabular}
    \caption{Number of integrand evaluations, precision of the integral, precision gain, and $N_{>4\sigma}$ for $N\approx 10000$ and $D=5$. Notations are the same as Table~\ref{tab:res5000-5}.}
    \label{tab:res10000-5}
\end{table}
%%%%%%%%%%%%%%%%%%%%%%%%%%%%%%%%%%%%%%%%%%
\begin{table}[htbp]
\footnotesize
    \centering
    \begin{tabular}{cc|cccccc}
\multicolumn{8}{c}{N$\approx$10000, D=8, $||\mathbf{c}||_1$=1.0}\\\hline\hline
\multicolumn{2}{c|}{Integrand family} & 1 & 2 & 3 & 4 & 5 & 6 \\\hline
\multicolumn{2}{c|}{$N$}&          10000(0) &          10000(0) &          10000(0) &          10000(0) &          10000(0) &          10000(0) \\
\multicolumn{2}{c|}{$N_{\textrm{train}}$}&           5000(0) &           5000(0) &           7000(0) &           5000(0) &           5000(0) &           3000(0) \\
\multicolumn{2}{c|}{$N_{\textrm{crxn}}$}&           5000(0) &           5000(0) &           3000(0) &           5000(0) &           5000(0) &          6948(27) \\
\hline
\multirow{4}{*}{$\sigma_I/|I|$}
& VEG & $7(5)\hspace{-0.3em}\times\hspace{-0.3em} 10^{-3}$ & $3.68(8)\hspace{-0.3em}\times\hspace{-0.3em} 10^{-4}$ & $1.00(1)\hspace{-0.3em}\times\hspace{-0.3em} 10^{-3}$ & $3.55(5)\hspace{-0.3em}\times\hspace{-0.3em} 10^{-4}$ & $5.35(5)\hspace{-0.3em}\times\hspace{-0.3em} 10^{-4}$ & $3.7(2)\hspace{-0.3em}\times\hspace{-0.3em} 10^{-3}$ \\
& MLP & $9(5)\hspace{-0.3em}\times\hspace{-0.3em} 10^{-5}$ & $6.0(1)\hspace{-0.3em}\times\hspace{-0.3em} 10^{-6}$ & $1.16(3)\hspace{-0.3em}\times\hspace{-0.3em} 10^{-4}$ & $5.7(1)\hspace{-0.3em}\times\hspace{-0.3em} 10^{-6}$ & $3.3(1)\hspace{-0.3em}\times\hspace{-0.3em} 10^{-5}$ & $5.0(5)\hspace{-0.3em}\times\hspace{-0.3em} 10^{-3}$ \\
& GBDT & $2(1)\hspace{-0.3em}\times\hspace{-0.3em} 10^{-3}$ & $1.11(1)\hspace{-0.3em}\times\hspace{-0.3em} 10^{-5}$ & $1.26(2)\hspace{-0.3em}\times\hspace{-0.3em} 10^{-3}$ & $1.11(1)\hspace{-0.3em}\times\hspace{-0.3em} 10^{-5}$ & $1.30(1)\hspace{-0.3em}\times\hspace{-0.3em} 10^{-4}$ & $3.3(1)\hspace{-0.3em}\times\hspace{-0.3em} 10^{-4}$ \\
& GP & $3(2)\hspace{-0.3em}\times\hspace{-0.3em} 10^{-5}$ & $4.3(2)\hspace{-0.3em}\times\hspace{-0.3em} 10^{-8}$ & $4.70(9)\hspace{-0.3em}\times\hspace{-0.3em} 10^{-6}$ & $3.7(1)\hspace{-0.3em}\times\hspace{-0.3em} 10^{-8}$ & $1.49(3)\hspace{-0.3em}\times\hspace{-0.3em} 10^{-4}$ & $1.9(2)\hspace{-0.3em}\times\hspace{-0.3em} 10^{-2}$ \\
\hline
\multirow{4}{*}{Gain}
& MLP &         60.6(5.4) &         61.4(1.2) &          8.80(23) &         62.8(1.2) &         16.69(51) &         0.791(35) \\
& GBDT &          4.84(38) &         33.42(90) &         0.802(21) &         32.06(52) &         4.129(89) &         10.95(57) \\
& GP &           359(28) &         9130(307) &        215.1(2.8) &         9649(203) &         3.650(98) &         0.230(14) \\
\hline
\multirow{1}{*}{$N_{>4\sigma}$}
& VEG &                -- &                -- &                -- &                -- &                -- &                 1 \\
\hline\hline\\
\multicolumn{8}{c}{N$\approx$10000, D=8, $||\mathbf{c}||_1$=3.0}\\\hline\hline
\multicolumn{2}{c|}{Integrand family} & 1 & 2 & 3 & 4 & 5 & 6 \\\hline
\multicolumn{2}{c|}{$N$}&          10000(0) &          10000(0) &          10000(0) &          10000(0) &          10000(0) &          10000(0) \\
\multicolumn{2}{c|}{$N_{\textrm{train}}$}&           5000(0) &           5000(0) &           7000(0) &           5000(0) &           5000(0) &           3000(0) \\
\multicolumn{2}{c|}{$N_{\textrm{crxn}}$}&           5000(0) &           5000(0) &           3000(0) &           5000(0) &           5000(0) &          6914(39) \\
\hline
\multirow{4}{*}{$\sigma_I/|I|$}
& VEG & $1.0(4)\hspace{-0.3em}\times\hspace{-0.3em} 10^{-2}$ & $5.56(7)\hspace{-0.3em}\times\hspace{-0.3em} 10^{-4}$ & $2.51(3)\hspace{-0.3em}\times\hspace{-0.3em} 10^{-3}$ & $5.54(6)\hspace{-0.3em}\times\hspace{-0.3em} 10^{-4}$ & $6.53(8)\hspace{-0.3em}\times\hspace{-0.3em} 10^{-4}$ & $3.8(2)\hspace{-0.3em}\times\hspace{-0.3em} 10^{-3}$ \\
& MLP & $1.6(6)\hspace{-0.3em}\times\hspace{-0.3em} 10^{-4}$ & $4.68(9)\hspace{-0.3em}\times\hspace{-0.3em} 10^{-5}$ & $3.7(4)\hspace{-0.3em}\times\hspace{-0.3em} 10^{-4}$ & $5.00(9)\hspace{-0.3em}\times\hspace{-0.3em} 10^{-5}$ & $1.02(4)\hspace{-0.3em}\times\hspace{-0.3em} 10^{-4}$ & $5.2(4)\hspace{-0.3em}\times\hspace{-0.3em} 10^{-3}$ \\
& GBDT & $7(3)\hspace{-0.3em}\times\hspace{-0.3em} 10^{-3}$ & $9.2(1)\hspace{-0.3em}\times\hspace{-0.3em} 10^{-5}$ & $4.12(6)\hspace{-0.3em}\times\hspace{-0.3em} 10^{-3}$ & $9.7(1)\hspace{-0.3em}\times\hspace{-0.3em} 10^{-5}$ & $3.98(5)\hspace{-0.3em}\times\hspace{-0.3em} 10^{-4}$ & $7.3(2)\hspace{-0.3em}\times\hspace{-0.3em} 10^{-4}$ \\
& GP & $6(2)\hspace{-0.3em}\times\hspace{-0.3em} 10^{-5}$ & $4.6(8)\hspace{-0.3em}\times\hspace{-0.3em} 10^{-7}$ & $1.77(3)\hspace{-0.3em}\times\hspace{-0.3em} 10^{-5}$ & $3.3(1)\hspace{-0.3em}\times\hspace{-0.3em} 10^{-7}$ & $4.6(1)\hspace{-0.3em}\times\hspace{-0.3em} 10^{-4}$ & $2.1(2)\hspace{-0.3em}\times\hspace{-0.3em} 10^{-2}$ \\
\hline
\multirow{4}{*}{Gain}
& MLP &         46.9(2.7) &         11.98(22) &          7.66(30) &         11.21(22) &          6.64(23) &         0.786(36) \\
& GBDT &         2.104(85) &          6.06(15) &         0.620(17) &         5.715(72) &         1.656(39) &          5.04(20) \\
& GP &        147.2(1.5) &          1592(89) &        142.3(1.2) &          1698(49) &         1.436(38) &         0.228(16) \\
\hline
\multirow{2}{*}{$N_{>4\sigma}$}
& VEG &                -- &                -- &                -- &                -- &                -- &                 3 \\
& GBDT &                -- &                -- &                 1 &                -- &                -- &                -- \\
\hline\hline\\
\multicolumn{8}{c}{N$\approx$10000, D=8, $||\mathbf{c}||_1$=8.0}\\\hline\hline
\multicolumn{2}{c|}{Integrand family} & 1 & 2 & 3 & 4 & 5 & 6 \\\hline
\multicolumn{2}{c|}{$N$}&          10000(0) &          10000(0) &          10000(0) &          10000(0) &          10000(0) &          10000(0) \\
\multicolumn{2}{c|}{$N_{\textrm{train}}$}&           5000(0) &           5000(0) &           7000(0) &           5000(0) &           5000(0) &           3000(0) \\
\multicolumn{2}{c|}{$N_{\textrm{crxn}}$}&           5000(0) &           5000(0) &           3000(0) &           5000(0) &           5000(0) &          7008(38) \\
\hline
\multirow{4}{*}{$\sigma_I/|I|$}
& VEG & $2.6(8)\hspace{-0.3em}\times\hspace{-0.3em} 10^{-2}$ & $7.7(1)\hspace{-0.3em}\times\hspace{-0.3em} 10^{-4}$ & $5.97(6)\hspace{-0.3em}\times\hspace{-0.3em} 10^{-3}$ & $8.8(1)\hspace{-0.3em}\times\hspace{-0.3em} 10^{-4}$ & $8.5(1)\hspace{-0.3em}\times\hspace{-0.3em} 10^{-4}$ & $5.7(2)\hspace{-0.3em}\times\hspace{-0.3em} 10^{-3}$ \\
& MLP & $7(3)\hspace{-0.3em}\times\hspace{-0.3em} 10^{-4}$ & $2.30(3)\hspace{-0.3em}\times\hspace{-0.3em} 10^{-4}$ & $7.5(5)\hspace{-0.3em}\times\hspace{-0.3em} 10^{-4}$ & $3.10(4)\hspace{-0.3em}\times\hspace{-0.3em} 10^{-4}$ & $2.86(8)\hspace{-0.3em}\times\hspace{-0.3em} 10^{-4}$ & $6.0(6)\hspace{-0.3em}\times\hspace{-0.3em} 10^{-3}$ \\
& GBDT & $3(2)\hspace{-0.3em}\times\hspace{-0.3em} 10^{-2}$ & $5.08(9)\hspace{-0.3em}\times\hspace{-0.3em} 10^{-4}$ & $1.35(2)\hspace{-0.3em}\times\hspace{-0.3em} 10^{-2}$ & $6.47(8)\hspace{-0.3em}\times\hspace{-0.3em} 10^{-4}$ & $1.16(1)\hspace{-0.3em}\times\hspace{-0.3em} 10^{-3}$ & $2.05(5)\hspace{-0.3em}\times\hspace{-0.3em} 10^{-3}$ \\
& GP & $3(1)\hspace{-0.3em}\times\hspace{-0.3em} 10^{-4}$ & $4.0(6)\hspace{-0.3em}\times\hspace{-0.3em} 10^{-5}$ & $6.3(2)\hspace{-0.3em}\times\hspace{-0.3em} 10^{-5}$ & $2.8(1)\hspace{-0.3em}\times\hspace{-0.3em} 10^{-6}$ & $1.36(3)\hspace{-0.3em}\times\hspace{-0.3em} 10^{-3}$ & $2.7(3)\hspace{-0.3em}\times\hspace{-0.3em} 10^{-2}$ \\
\hline
\multirow{4}{*}{Gain}
& MLP &         51.3(1.7) &         3.383(78) &          8.81(44) &         2.851(50) &         3.075(89) &         1.126(68) \\
& GBDT &         1.252(32) &         1.546(47) &         0.453(10) &         1.368(28) &         0.741(17) &         2.746(84) \\
& GP &        132.3(2.0) &         39.7(5.0) &         97.2(2.9) &        319.5(7.7) &         0.638(15) &         0.269(20) \\
\hline
\multirow{2}{*}{$N_{>4\sigma}$}
& VEG &                -- &                -- &                -- &                -- &                -- &                 1 \\
& MLP &                -- &                 1 &                 3 &                -- &                -- &                -- \\
\hline\hline
    \end{tabular}
    \caption{Number of integrand evaluations, precision of the integral, precision gain, and $N_{>4\sigma}$ for $N\approx 10000$ and $D=8$. Notations are the same as Table~\ref{tab:res5000-5}.}
    \label{tab:res10000-8}
\end{table}
%%%%%%%%%%%%%%%%%%%%%%%%%%%%%%%%%%%%%%%%%%
\begin{table}[htbp]
\footnotesize
    \centering
    \begin{tabular}{cc|cccccc}
\multicolumn{8}{c}{N$\approx$10000, D=10, $||\mathbf{c}||_1$=1.0}\\\hline\hline
\multicolumn{2}{c|}{Integrand family} & 1 & 2 & 3 & 4 & 5 & 6 \\\hline
\multicolumn{2}{c|}{$N$}&          10000(0) &          10000(0) &          10000(0) &          10000(0) &          10000(0) &          10000(0) \\
\multicolumn{2}{c|}{$N_{\textrm{train}}$}&           5000(0) &           5000(0) &           7000(0) &           5000(0) &           5000(0) &           3000(0) \\
\multicolumn{2}{c|}{$N_{\textrm{crxn}}$}&           5000(0) &           5000(0) &           3000(0) &           5000(0) &           5000(0) &           7000(0) \\
\hline
\multirow{4}{*}{$\sigma_I/|I|$}
& VEG & $5(3)\hspace{-0.3em}\times\hspace{-0.3em} 10^{-3}$ & $3.4(1)\hspace{-0.3em}\times\hspace{-0.3em} 10^{-4}$ & $1.13(1)\hspace{-0.3em}\times\hspace{-0.3em} 10^{-3}$ & $3.21(5)\hspace{-0.3em}\times\hspace{-0.3em} 10^{-4}$ & $5.90(5)\hspace{-0.3em}\times\hspace{-0.3em} 10^{-4}$ & $3.8(3)\hspace{-0.3em}\times\hspace{-0.3em} 10^{-3}$ \\
& MLP & $9(6)\hspace{-0.3em}\times\hspace{-0.3em} 10^{-5}$ & $5.9(1)\hspace{-0.3em}\times\hspace{-0.3em} 10^{-6}$ & $1.42(4)\hspace{-0.3em}\times\hspace{-0.3em} 10^{-4}$ & $5.6(1)\hspace{-0.3em}\times\hspace{-0.3em} 10^{-6}$ & $4.8(2)\hspace{-0.3em}\times\hspace{-0.3em} 10^{-5}$ & $6(1)\hspace{-0.3em}\times\hspace{-0.3em} 10^{-3}$ \\
& GBDT & $3(2)\hspace{-0.3em}\times\hspace{-0.3em} 10^{-3}$ & $1.05(1)\hspace{-0.3em}\times\hspace{-0.3em} 10^{-5}$ & $1.91(2)\hspace{-0.3em}\times\hspace{-0.3em} 10^{-3}$ & $1.04(1)\hspace{-0.3em}\times\hspace{-0.3em} 10^{-5}$ & $1.51(1)\hspace{-0.3em}\times\hspace{-0.3em} 10^{-4}$ & $3.7(1)\hspace{-0.3em}\times\hspace{-0.3em} 10^{-4}$ \\
& GP & $4(3)\hspace{-0.3em}\times\hspace{-0.3em} 10^{-5}$ & $3.9(2)\hspace{-0.3em}\times\hspace{-0.3em} 10^{-8}$ & $6.3(1)\hspace{-0.3em}\times\hspace{-0.3em} 10^{-6}$ & $3.40(8)\hspace{-0.3em}\times\hspace{-0.3em} 10^{-8}$ & $1.59(3)\hspace{-0.3em}\times\hspace{-0.3em} 10^{-4}$ & $2.1(2)\hspace{-0.3em}\times\hspace{-0.3em} 10^{-2}$ \\
\hline
\multirow{4}{*}{Gain}
& MLP &         59.3(5.0) &         57.6(1.1) &          8.16(21) &         57.39(95) &         12.95(51) &         0.696(33) \\
& GBDT &          4.52(40) &         32.5(1.1) &         0.598(13) &         30.96(53) &         3.918(71) &          9.84(58) \\
& GP &           339(27) &         9092(218) &        180.5(2.0) &         9535(118) &          3.78(10) &         0.218(11) \\
\hline
\multirow{1}{*}{$N_{>4\sigma}$}
& -- &                -- &                -- &                -- &                -- &                -- &                -- \\
\hline\hline\\
\multicolumn{8}{c}{N$\approx$10000, D=10, $||\mathbf{c}||_1$=3.0}\\\hline\hline
\multicolumn{2}{c|}{Integrand family} & 1 & 2 & 3 & 4 & 5 & 6 \\\hline
\multicolumn{2}{c|}{$N$}&          10000(0) &          10000(0) &          10000(0) &          10000(0) &          10000(0) &          10000(0) \\
\multicolumn{2}{c|}{$N_{\textrm{train}}$}&           5000(0) &           5000(0) &           7000(0) &           5000(0) &           5000(0) &           3000(0) \\
\multicolumn{2}{c|}{$N_{\textrm{crxn}}$}&           5000(0) &           5000(0) &           3000(0) &           5000(0) &           5000(0) &           7000(0) \\
\hline
\multirow{4}{*}{$\sigma_I/|I|$}
& VEG & $9(4)\hspace{-0.3em}\times\hspace{-0.3em} 10^{-3}$ & $5.97(8)\hspace{-0.3em}\times\hspace{-0.3em} 10^{-4}$ & $2.83(3)\hspace{-0.3em}\times\hspace{-0.3em} 10^{-3}$ & $5.89(5)\hspace{-0.3em}\times\hspace{-0.3em} 10^{-4}$ & $7.16(7)\hspace{-0.3em}\times\hspace{-0.3em} 10^{-4}$ & $4.2(3)\hspace{-0.3em}\times\hspace{-0.3em} 10^{-3}$ \\
& MLP & $1.8(7)\hspace{-0.3em}\times\hspace{-0.3em} 10^{-4}$ & $4.9(1)\hspace{-0.3em}\times\hspace{-0.3em} 10^{-5}$ & $4.5(2)\hspace{-0.3em}\times\hspace{-0.3em} 10^{-4}$ & $5.07(8)\hspace{-0.3em}\times\hspace{-0.3em} 10^{-5}$ & $1.45(5)\hspace{-0.3em}\times\hspace{-0.3em} 10^{-4}$ & $6.9(9)\hspace{-0.3em}\times\hspace{-0.3em} 10^{-3}$ \\
& GBDT & $9(4)\hspace{-0.3em}\times\hspace{-0.3em} 10^{-3}$ & $8.9(1)\hspace{-0.3em}\times\hspace{-0.3em} 10^{-5}$ & $6.76(8)\hspace{-0.3em}\times\hspace{-0.3em} 10^{-3}$ & $9.2(1)\hspace{-0.3em}\times\hspace{-0.3em} 10^{-5}$ & $4.61(5)\hspace{-0.3em}\times\hspace{-0.3em} 10^{-4}$ & $8.7(2)\hspace{-0.3em}\times\hspace{-0.3em} 10^{-4}$ \\
& GP & $8(3)\hspace{-0.3em}\times\hspace{-0.3em} 10^{-5}$ & $5(1)\hspace{-0.3em}\times\hspace{-0.3em} 10^{-7}$ & $2.45(4)\hspace{-0.3em}\times\hspace{-0.3em} 10^{-5}$ & $3.04(7)\hspace{-0.3em}\times\hspace{-0.3em} 10^{-7}$ & $4.9(1)\hspace{-0.3em}\times\hspace{-0.3em} 10^{-4}$ & $2.2(2)\hspace{-0.3em}\times\hspace{-0.3em} 10^{-2}$ \\
\hline
\multirow{4}{*}{Gain}
& MLP &         39.9(1.9) &         12.32(27) &          6.57(20) &         11.70(19) &          5.14(19) &         0.722(41) \\
& GBDT &         1.832(71) &          6.70(14) &        0.4227(74) &         6.373(81) &         1.561(28) &          4.65(26) \\
& GP &        118.1(1.1) &         1679(111) &        116.3(1.1) &          1976(47) &         1.488(40) &         0.227(13) \\
\hline
\multirow{1}{*}{$N_{>4\sigma}$}
& VEG &                -- &                -- &                -- &                -- &                -- &                 1 \\
\hline\hline\\
\multicolumn{8}{c}{N$\approx$10000, D=10, $||\mathbf{c}||_1$=8.0}\\\hline\hline
\multicolumn{2}{c|}{Integrand family} & 1 & 2 & 3 & 4 & 5 & 6 \\\hline
\multicolumn{2}{c|}{$N$}&          10000(0) &          10000(0) &          10000(0) &          10000(0) &          10000(0) &          10000(0) \\
\multicolumn{2}{c|}{$N_{\textrm{train}}$}&           5000(0) &           5000(0) &           7000(0) &           5000(0) &           5000(0) &           3000(0) \\
\multicolumn{2}{c|}{$N_{\textrm{crxn}}$}&           5000(0) &           5000(0) &           3000(0) &           5000(0) &           5000(0) &           7000(0) \\
\hline
\multirow{4}{*}{$\sigma_I/|I|$}
& VEG & $2(1)\hspace{-0.3em}\times\hspace{-0.3em} 10^{-2}$ & $8.2(1)\hspace{-0.3em}\times\hspace{-0.3em} 10^{-4}$ & $7.06(9)\hspace{-0.3em}\times\hspace{-0.3em} 10^{-3}$ & $8.70(9)\hspace{-0.3em}\times\hspace{-0.3em} 10^{-4}$ & $9.21(9)\hspace{-0.3em}\times\hspace{-0.3em} 10^{-4}$ & $5.5(3)\hspace{-0.3em}\times\hspace{-0.3em} 10^{-3}$ \\
& MLP & $7(4)\hspace{-0.3em}\times\hspace{-0.3em} 10^{-4}$ & $2.59(3)\hspace{-0.3em}\times\hspace{-0.3em} 10^{-4}$ & $1.0(1)\hspace{-0.3em}\times\hspace{-0.3em} 10^{-3}$ & $3.31(5)\hspace{-0.3em}\times\hspace{-0.3em} 10^{-4}$ & $4.1(2)\hspace{-0.3em}\times\hspace{-0.3em} 10^{-4}$ & $7.6(8)\hspace{-0.3em}\times\hspace{-0.3em} 10^{-3}$ \\
& GBDT & $1(1)\hspace{-0.3em}\times\hspace{-0.3em} 10^{-1}$ & $5.30(8)\hspace{-0.3em}\times\hspace{-0.3em} 10^{-4}$ & $2.43(6)\hspace{-0.3em}\times\hspace{-0.3em} 10^{-2}$ & $6.37(7)\hspace{-0.3em}\times\hspace{-0.3em} 10^{-4}$ & $1.34(1)\hspace{-0.3em}\times\hspace{-0.3em} 10^{-3}$ & $2.44(5)\hspace{-0.3em}\times\hspace{-0.3em} 10^{-3}$ \\
& GP & $3(1)\hspace{-0.3em}\times\hspace{-0.3em} 10^{-4}$ & $4.5(8)\hspace{-0.3em}\times\hspace{-0.3em} 10^{-5}$ & $1.30(6)\hspace{-0.3em}\times\hspace{-0.3em} 10^{-4}$ & $2.39(6)\hspace{-0.3em}\times\hspace{-0.3em} 10^{-6}$ & $1.42(2)\hspace{-0.3em}\times\hspace{-0.3em} 10^{-3}$ & $2.9(3)\hspace{-0.3em}\times\hspace{-0.3em} 10^{-2}$ \\
\hline
\multirow{4}{*}{Gain}
& MLP &         42.69(95) &         3.181(47) &          7.67(36) &         2.645(44) &          2.39(10) &         0.872(56) \\
& GBDT &         1.100(26) &         1.570(43) &        0.2983(86) &         1.371(19) &         0.690(12) &         2.198(93) \\
& GP &         88.7(2.6) &         35.6(4.0) &         56.9(2.1) &        372.1(9.0) &         0.657(15) &         0.236(17) \\
\hline
\multirow{3}{*}{$N_{>4\sigma}$}
& VEG &                -- &                -- &                -- &                -- &                -- &                 2 \\
& MLP &                -- &                -- &                 7 &                -- &                -- &                -- \\
& GP &                -- &                -- &                 1 &                -- &                -- &                -- \\
\hline\hline
    \end{tabular}
    \caption{Number of integrand evaluations, precision of the integral, precision gain, and $N_{>4\sigma}$ for $N\approx 10000$ and $D=8$. Notations are the same as Table~\ref{tab:res5000-5}.}
    \label{tab:res10000-10}
\end{table}

\bibliographystyle{apsrev4-1}
\bibliography{ref}

\end{document}